\documentclass[10pt,english]{elsarticle}
\usepackage[T1]{fontenc}
\usepackage[latin9]{inputenc}
\usepackage[a4paper]{geometry}
\usepackage[active]{srcltx}
\usepackage{textcomp}
\usepackage{amsmath}
\usepackage{amssymb}
\usepackage{graphicx}
\usepackage{wasysym}

\makeatletter

\newcommand*\LyXThinSpace{\,\hspace{0pt}}

\makeatother

\usepackage{babel}
\begin{document}

\begin{frontmatter}{}

\title{Finite size effects around pseudo-transition in one-dimensional models
with nearest neighbor interaction}

\author{Onofre Rojas}

\address{Department of Physics, Federal University of Lavras, 37200-900 Lavras-MG,
Brazil}
\begin{abstract}
Recently gigantic peaks in thermodynamic response functions have been
observed at finite temperature for one-dimensional models with short-range
coupling, closely resembling a second-order phase transition. Thus,
we will analyze the finite temperature pseudo-transition property
observed in some one-dimensional models and its relationship with
finite size effect. In particular, we consider two chain models to
study the finite size effects; these are the Ising-Heisenberg tetrahedral
chain and an Ising-Heisenberg-type ladder model. Although the anomalous
peaks of these one-dimensional models have already been studied in
the thermodynamic limit, here we will discuss the finite size effects
of the chain and why the peaks do not diverge in the thermodynamic
limit. So, we discuss the dependence of the finite size effects, for
moderately and sufficiently large systems, in which the specific heat
and magnetic susceptibility exhibit peculiar rounded towering peaks
for a given temperature. This behavior is quite similar to a continuous
phase transition, but there is no singularity. For moderately large
systems, the peaks narrow and increase in height as the number of
unit cells is increased, and the location of peak shifts slightly.
Hence, one can naively induce that the sharp peak should lead to a
divergence in the thermodynamic limit. However, for a rather large
system, the height of a peak goes asymptotically to a finite value.
Our result rigorously confirms the dependence of the peak height with
the number of unit cells at the pseudo-critical temperature. We also
provide an alternative empirical function that satisfactorily fits
specific heat and magnetic susceptibility at pseudo-critical temperature.
Certainly, our result is crucial to understand the finite size correction
behavior in quantum spin models, which in general are only numerically
tractable within the framework of the finite size analysis.
\end{abstract}

\end{frontmatter}{}

\section{Introduction}

Although one-dimensional models usually do not describe the phase
transition at finite temperature, there are some unusual one-dimensional
models with a short-range coupling exhibiting a first-order phase
transition at finite temperature. Recently, Sarkanych et al.\citep{sarkanych}
proposed a one-dimensional Potts model with an additional energy degeneracy,
which contributes to entropy but not the interaction energy. This
extra degeneracy generates a first-order phase transition. The Kittel
or zipper model\citep{kittel}, is a typical simple model whose transfer
matrix is finite-dimensional. The zipper model constraint leads to
infinite potential, so the free energy becomes a non-analytic function,
exhibiting a first-order phase transition. Another model that we can
mention is that considered by Chui-Weeks\citep{chui}, which was proposed
to study the solid-on-solid for the surface growth, whose transfer
matrix has an infinite dimension, but still analytically tractable.
Imposing the impenetrable condition to subtract, the model shows the
existence of the phase transition. The Dauxois-Peyrard\citep{dauxois}
is another model with an infinite-dimensional transfer matrix, where
some evidence of phase transition has been found. In summary, all
those models break the Perron-Frobenius theorem\citep{Ninio} since
free energy becomes non-analytical at a critical temperature, or equivalently
some elements of the transfer matrix become null (corresponding to
an infinite energy).

Previously in the literature, an anomalous property in one-dimensional
models with short-range coupling was observed, where an abrupt continuous
change comes out in the first derivative of free energy at finite
temperature, somewhat similar to the first-order phase transition.
Whereas for the second derivative of free energy, an intense giant
peak comes into sight, although there is no discontinuity or divergence,
which resembles a second-order phase transition. Earlier, in 2011
Timonin\citep{Timonin} called this phenomenon as \textquotedblleft pseudo-transition\textquotedblright{}
while studying the spin ice in a field, to indicate a sudden change
in the first derivative of free energy, and a vigorous peak in the
second derivative of free energy, although there is no discontinuity
or divergence, in physical quantities. It is worth noting that the
term pseudo-transition is generally used to study finite size lattice
systems. Thus, it is possible to observe peaks that increase with
the number of lattice sites, by performing the extrapolation techniques,
it is possible to conclude the lattice exhibits a real phase transition
in the thermodynamic limit. The results found here are quite similar,
for moderately large chain sizes the peak increases proportionally
to the chain size, although for chains with a sufficiently large number
of unit cells it saturates to a finite value. This fact justifies
the name of pseudo-transition.

Lately, this unusual property was discussed in the following recent
works. A double-tetrahedral chain of localized Ising spins and mobile
electrons show a strong thermal excitation that resembles a first-order
phase transition\citep{Galisova,galisova17}. In the frustrated spin-1/2,
Ising-Heisenberg's three-leg tube exhibited a pseudo-transition\citep{strk-cav}.
In reference \citep{on-strk}, a similar property was observed when
studying thermal entanglement, whose specific heat was reported with
a sharp peak on the spin-1/2 Ising-Heisenberg ladder with alternating
Ising and Heisenberg inter-leg couplings. This anomalous property
was still observed in the spin-1/2 Ising diamond chain in the neighboring
of the pseudo-transition\citep{psd-Ising}. Besides, we also discuss
additional properties and further investigations into this peculiar
property\citep{pseudo}. As well as that considered distant correlation
functions for a spin-1/2 Ising-XYZ diamond chain\citep{Isaac}. An
alternative proposal to identify the pseudo-transition was analyzed
in the phase boundary residual entropy representation and its finite
temperature pseudo-transition relation in one-dimensional models\citep{ph-bd}.
The universality and pseudo-critical exponents of one-dimensional
models were also discussed around the pseudo-transition\citep{unv-cr-exp}.

It is valuable to note that the pseudo-transition does not violate
the Perron-Frobenius theorem\citep{Ninio}, and guarantees the largest
eigenvalues of the transfer matrix are non-degenerate, so free energy
becomes an analytical function.  Equivalently, some elements (Boltzmann
factor) of the transfer matrix become just a tiny amount compared
to other elements, or the corresponding energy becomes large enough
but finite compared to ground state energy. These types of properties
appear more frequently in decorated models\citep{Dec-trnsf}. By using
somewhat different perspectives, similar anomalous properties have
also been discussed in the references \citep{Hutak,Weiguo}.

There is a natural question, how to find a pseudo-critical temperature
in quantum spin systems? Such as one-dimensional quantum spin models,
quantum spin ladder models\citep{Tsai}, quantum spin tube models\citep{R. Chen}
or one-dimensional Hubbard models\citep{Zhao,Zhang,Ding,Shi}. In
this sense, the present work concomitant the proposed technique in
reference \citep{ph-bd} would useful to elucidate some evidence of
pseudo-transition in finite size quantum spin chain. Since most of
the quantum spin chain models are only numerically tractable. 

We organized this article as follows: In the second section, we present
the transfer matrix technique around the pseudo-critical temperature
and the finite size   effects. In section 3, we discuss Ising-Heisenberg's
tetrahedral chain thermodynamics. In sec 4, we apply for a decorated
Ising Heisenberg model in the pseudo-transition vicinity for a finite
length chain. We also propose a simple empirical function that fits
precisely with exact results. Finally, in section 5, we give our conclusions
and perspectives.

\section{Transfer matrix}

The transfer matrix technique in statistical physics was introduced
in 1941 by H. Kramers and G. Wannier\citep{kramers}. Since then,
several lattice models partition functions would be obtained using
this technique, where basically, the partition function is written
as a sum of all possible micro-states. It also includes an additional
summation of each energy level contribution of the system within each
micro-state. A somewhat general transfer matrix can always be expressed
as a symmetric matrix 

\begin{equation}
\mathbf{V}=\left[\begin{array}{cccc}
v_{1,1} & v_{1,2} & \cdots & v_{1,n}\\
v_{1,2} & v_{2,2} & \cdots & v_{2,n}\\
\vdots & \vdots & \ddots & \vdots\\
v_{1,n} & v_{2,n} & \cdots & v_{n,n}
\end{array}\right],
\end{equation}
where the elements are assumed to be $v_{i,j}>0$.

Therefore, the partition function becomes 

\begin{equation}
\mathcal{Z}_{N}={\rm tr}\left(\prod_{k=1}^{N}\mathbf{V}_{k}\right)={\rm tr}\left(\mathbf{V}^{N}\right).
\end{equation}
In order to obtain the partition function, we must first diagonalize
the transfer matrix $\mathbf{V}$, assuming whose eigenvalues are
denoted by $\{\lambda_{r}\}$. So we can express the partition function
as follows 

\begin{equation}
\mathcal{Z}_{N}=\sum_{r=1}^{n}\lambda_{r}^{N}.\label{eq:part-func}
\end{equation}
According to the Perron-Frobenius theorem\citep{Ninio,ky lin}, there
is a largest non-degenerate eigenvalue $\lambda_{1}$ of the transfer
matrix that satisfy $\lambda_{1}>\lambda_{r}$ with $r=2,3,\dots,n$.
Here $N$ stands for the number of sites or unit cells, but it is
common in practice to assume $N\rightarrow\infty$. However here we
focus on exploring the finite size corrections of one-dimensional
models. 

Using the partition function, we can write free energy per site, as
in many textbooks provided by 
\begin{equation}
f^{(N)}=-\frac{1}{\beta N}\ln\left(\mathcal{Z}_{N}\right)=-\frac{1}{N\beta}\ln\left\{ \lambda_{1}^{N}\left[1+\sum_{r=2}^{n}\left(\frac{\lambda_{r}}{\lambda_{1}}\right)^{N}\right]\right\} .\label{eq:free_Nexp}
\end{equation}
Therefore, the free energy\eqref{eq:free_Nexp} can be expressed as
follows

\begin{equation}
f^{(N)}=-\frac{1}{\beta}\ln\left(\lambda_{1}\right)-\frac{1}{N\beta}\ln\left[1+\sum_{r=2}^{n}\left(\frac{\lambda_{r}}{\lambda_{1}}\right)^{N}\right],\label{eq:free-N}
\end{equation}
note that each $\frac{\lambda_{r}}{\lambda_{1}}<1$, with $r\geqslant2$.
Then $\left(\frac{\lambda_{r}}{\lambda_{1}}\right)^{N}\rightarrow0$
when $N\rightarrow\infty$, thus the free energy in thermodynamic
limit reduces to 

\begin{equation}
f=-\frac{1}{\beta}\ln\left(\lambda_{1}\right).\label{eq:f_e}
\end{equation}

In principle, the largest eigenvalue may becomes degenerate $\lambda_{1}=\lambda_{2}$,
for some specific control parameters at finite temperature, this would
mean the existence of a discontinuous phase transition for a given
control parameter. Then undoubtedly, the Perron-Frobenius theorem\citep{Ninio}
must be broken. Alternatively, this analysis will be considered in
detail below.

\subsection{Finite size effects for \textquotedbl quasi-degenerate\textquotedbl{}
largest eigenvalues }

Now let us analyze the eigenvalues of transfer matrix using a slightly
different perspective. For that purpose, we assume that the Perron-Frobenius
theorem\citep{Ninio} is never broken. Therefore, let as back to eq.\eqref{eq:free-N}
and assume that $\lambda_{2}\rightarrow\lambda_{1}$ (but still $\lambda_{2}<\lambda_{1}$)
for some particular control parameter. This means that the largest
eigenvalue becomes almost degenerate\textsl{ (quasi-degenerate)} but
still satisfying the Perron-Frobenius theorem. Taking this fact into
account, the free energy provided by \eqref{eq:free-N}, can be written
as 

\begin{equation}
f^{(N)}=f-\frac{1}{N\beta}\ln\left[1+\left(\frac{\lambda_{2}}{\lambda_{1}}\right)^{N}+\sum_{r=3}^{n}\left(\frac{\lambda_{r}}{\lambda_{1}}\right)^{N}\right].\label{eq:free-f-N}
\end{equation}

At this limit, the last term within the logarithm of \eqref{eq:free-f-N}
could be  properly neglected, for moderately large $N$.

On the other hand, let us denote conveniently a couple of largest
eigenvalues by
\begin{equation}
\lambda_{1,2}={\rm e}^{-\beta\varepsilon_{0}}\left(1\pm\frac{\zeta}{2}\right),\label{eq:mudanc}
\end{equation}
here we assume $\varepsilon_{0}$ as the lowest energy, and $\zeta$
is a parameter that depends of temperature.

When $\zeta\rightarrow0^{+}$, we have $\lambda_{1,2}\rightarrow{\rm e}^{-\beta\varepsilon_{0}}$.
This result induces us to believe that there is a degenerate state.
Still, according to Perron-Frobenius's theorem\citep{Ninio}, there
is no degeneracy, so we name that $\lambda_{1}$ and $\lambda_{2}$
are \textquotedbl quasi-degenerate\textquotedbl{} for a given temperature. 

Thereby, we can rewrite the free energy \eqref{eq:free-f-N} as a
function on $\zeta$ and ignoring the last term, we have
\begin{equation}
f^{(N)}=f-\tfrac{1}{N\beta}\ln\left[1+\left(\frac{1-\zeta/2}{1+\zeta/2}\right)^{N}\right],\label{eq:free-energ-N-1}
\end{equation}
and here $f$ denotes the free energy in the thermodynamic limit given
by \eqref{eq:f_e}.

By using the approximate result $\ln(1-\frac{\zeta}{2})\approx-\frac{\zeta}{2}$,
we can achieve to the following relation, 
\begin{equation}
\left(\frac{1-\zeta/2}{1+\zeta/2}\right)^{N}\approx{\rm e}^{-\zeta N},
\end{equation}
where we assume a finite positive $N$.

Consequently, the free energy for a finite chain in the vicinity of
quasi-degenerate transfer matrix eigenvalues, merely becomes 
\begin{equation}
f^{(N)}\approx f-\tfrac{T}{N}\ln\left(1+{\rm e}^{-\zeta N}\right).\label{eq:free-energ-N-2}
\end{equation}
This result describes the exact result accurately in the case when
the eigenvalues satisfy $\lambda_{2}\rightarrow\lambda_{1}$ or equivalently
$\zeta\rightarrow0^{+}$. 

On the other hand, by using the correlation length relation $\xi=\left[\ln\left(\frac{\lambda_{1}}{\lambda_{2}}\right)\right]^{-1}$,
we can express \eqref{eq:free-f-N} as follows
\begin{equation}
f^{(N)}=f-\tfrac{T}{N}\ln\left(1+{\rm e}^{-N/\xi}\right).\label{eq:f-corrl}
\end{equation}

Comparing the free energy \eqref{eq:free-energ-N-2} and \eqref{eq:f-corrl}
, we have the following relation $\zeta=\xi^{-1}$. However, this
result is only valid when $\zeta\rightarrow0$. 

\subsection{Thermodynamic limit close enough to pseudo-critical temperature }

For a given temperature, we have $\zeta\rightarrow0$ and $\lambda_{1,2}\rightarrow{\rm e}^{-\varepsilon_{0}/T_{p}}$,
which we call pseudo-critical temperature $T_{p}$. Therefore, we
can express the free energy in thermodynamic limit$f$ around pseudo-critical
temperature, in terms of the variables $\varepsilon_{0}$ and $\zeta$
according to the eqs.\eqref{eq:mudanc}, thus we have

\begin{alignat}{1}
f= & -T\ln({\rm e}^{-\beta\varepsilon_{0}})-T\ln\left(1+\frac{\zeta}{2}\right),\nonumber \\
f= & \varepsilon_{0}-\frac{T\zeta}{2}.\label{eq:free-e-crit}
\end{alignat}

It is worthy to note that $\zeta$ depends on the temperature, whose
minimum occurs for a given temperature called the pseudo-critical
temperature $T_{p}$. 

Using the free energy provided by \eqref{eq:free-e-crit}, we express
the entropy

\begin{equation}
\mathcal{S}=-\frac{\partial f}{\partial T}=\frac{\zeta}{2}+\frac{T}{2}\zeta_{T},\label{eq:S_e}
\end{equation}
where $\zeta_{T}=\frac{\partial\zeta}{\partial T}$ .

Likewise, the specific heat can be obtained as follows 

\begin{equation}
C=-T\frac{\partial^{2}f}{\partial T^{2}}=\frac{T}{2}\left(2\zeta_{T}+T\zeta_{T^{2}}\right),\label{eq:C_e}
\end{equation}
here we define $\zeta_{T^{2}}=\frac{\partial^{2}\zeta}{\partial T^{2}}$. 

Since, the magnetization is provided deriving the free energy respect
to the magnetic field, which results in

\begin{equation}
M=-\frac{\partial f}{\partial h}=T\zeta_{h},\label{eq:M_e}
\end{equation}
with $\zeta_{h}=\frac{\partial\zeta}{\partial h}$.

Analogously, using the free energy given in \eqref{eq:free-e-crit},
we write down the magnetic susceptibility as follows

\begin{equation}
\chi=-\frac{\partial^{2}f}{\partial h^{2}}=\frac{T\zeta_{h^{2}}}{2},\label{eq:chi_e}
\end{equation}
where the second derivative is denoted by $\zeta_{h^{2}}=\frac{\partial^{2}\zeta}{\partial h^{2}}$. 

This is a crucial point for exploring more cumbersome models, such
as quantum Heisenberg spin chain models, in which most of them are
only numerically tractable.

\subsection{Finite size physical quantities close enough to the pseudo-transition}

Our next task is to study the thermodynamics around the pseudo-transition
region, taking into account the finite size effects. Below, we can
obtain some physical quantities for a finite length chain around the
quasi-degenerate or pseudo-transition region, where correlation length
can approximately given by $\xi=\zeta^{-1}$, and using the free energy
given by \eqref{eq:free-energ-N-2}, we can obtain the quantities
below as a function of $N$.

The entropy for a finite size chain has the following form,
\begin{equation}
\mathcal{S}^{(N)}=\mathcal{S}+\frac{1}{N}\ln\left(1+{\rm e}^{-\zeta N}\right)-\frac{T\zeta_{T}}{1+{\rm e}^{\zeta N}},\label{eq:S_N}
\end{equation}
 where $\mathcal{S}$ means entropy in the thermodynamic limit, while
$\mathcal{S}^{(N)}$ denotes finite size chain entropy per unit cell.
In this region the entropy $\mathcal{S}$ describes a strong increase
at $T_{p}$, as we can observe in figs. \ref{fig:(a)-Specific-heat}a
and \ref{fig:Specific-heat6}a, as well as in references \citep{Galisova,strk-cav,pseudo,Isaac,ph-bd,unv-cr-exp}.

Another quantity is the specific heat for a finite size chain, which
can be obtained straightforwardly from eq.\eqref{eq:S_N}. Hence,
around the pseudo-critical temperature it becomes
\begin{equation}
C^{(N)}=C-\frac{T\left(2\zeta_{T}+T\zeta_{T^{2}}\right)}{\left(1+{\rm e}^{\zeta N}\right)}+\frac{T^{2}\zeta_{T}^{2}N{\rm e}^{\zeta N}}{\left(1+{\rm e}^{\zeta N}\right)^{2}},
\end{equation}
here $C$ denotes the specific heat in thermodynamic limit given by
\eqref{eq:C_e}. It is evident that $\zeta>0$, and when $\zeta\rightarrow0$
means that $\zeta$ has its minimum, so it is reasonable to assume
that around the pseudo-critical temperature we must have $\zeta_{T}\rightarrow0$.
Then $C^{(N)}$ reduces to the following expression
\begin{equation}
C^{(N)}=C-\frac{2C}{1+{\rm e}^{\zeta N}}=\left(\frac{{\rm e}^{\zeta N}-1}{1+{\rm e}^{\zeta N}}\right)C=\tanh\left(\frac{\zeta N}{2}\right)C.\label{eq:CN-ex}
\end{equation}
Here, the specific heat $C$ should show a strong sharp peak\citep{unv-cr-exp}
at $T_{p}$, as illustrated in figs. \ref{fig:(a)-Specific-heat}b
and \ref{fig:Specific-heat6}b.

Next, we express the magnetization for a given a finite size chain
\begin{equation}
M^{(N)}=M-\frac{T\zeta_{h}}{\left(1+{\rm e}^{\zeta N}\right)},
\end{equation}
where $M$ is given by \eqref{eq:M_e}, which means the magnetization
in the thermodynamic limit. Likewise, magnetization must report a
strong change at $T_{p}$.

Straightforwardly, we can also obtain the magnetic susceptibility
for a finite size   chain, 
\begin{equation}
\chi^{(N)}=\chi-\frac{T\zeta_{h^{2}}}{\left(1+{\rm e}^{\zeta N}\right)}+\frac{T\zeta_{h}^{2}N{\rm e}^{\zeta N}}{\left(1+{\rm e}^{\zeta N}\right)^{2}}.
\end{equation}

In a similar way to previous case, here we have $\zeta_{h}=\frac{\partial\zeta}{\partial h}\rightarrow0$
in the vicinity of the pseudo-critical transition. And we can surely
also ignore the second term, because of $\zeta_{h}^{2}<\zeta_{h}$.

Therefore, the magnetic susceptibility simply reduces to 
\begin{equation}
\chi^{(N)}=\chi-\frac{2\chi}{1+{\rm e}^{\zeta N}},
\end{equation}
here $\chi$ is given by \eqref{eq:chi_e}, which corresponds to magnetic
susceptibility in the thermodynamic limit. Simplifying the magnetic
susceptibility $\chi^{(N)}$, we achieve to the following simple expression,
\begin{equation}
\chi^{(N)}=\tanh\left(\frac{\zeta N}{2}\right)\chi.\label{eq:CN-tan}
\end{equation}
Magnetic susceptibility might also illustrate a strong acute peak\citep{unv-cr-exp}
close to $T_{p}$, as illustrated in figs. \ref{fig:(a)-Specific-heat}c
and \ref{fig:Specific-heat6}c.

To check the validity of the results \eqref{eq:CN-ex} and \eqref{eq:CN-tan},
we will apply for a couple of models in the next sections.

\subsection{Finite size correction around $T_{p}$ for moderately large $N$}

For moderately large $N$ and small $\zeta$ (near pseudo-critical
temperature), we can still simplify the eq.\eqref{eq:CN-ex}, which
becomes 
\begin{equation}
C^{(N)}\thickapprox\frac{C\,\zeta}{2}N.\label{eq:C_sclng}
\end{equation}

The specific heat \eqref{eq:C_sclng} indicate that the pseudo-critical
peak increases proportionally to $N$, this one resembles the finite
size corrections. Consequently, one might naively conclude when $N\rightarrow\infty$
leads to divergence, indicating a phase transition. However, it is
worth noting that this result fails for considerable large $N$, see
eq.\eqref{eq:CN-ex}. 

Using similar reasoning, for moderately large $N$ and small $\zeta$,
the result given in eq.\eqref{eq:CN-tan} leads to 
\begin{equation}
\chi^{(N)}\thickapprox\frac{\chi\,\zeta}{2}N.\label{eq:X_sclng}
\end{equation}
So the magnetic susceptibility increases proportionally to $N$ which
again resembles the finite size correction behavior, although, for
$N$ sufficiently large, this result also fails accordingly \eqref{eq:CN-tan}.

This analysis may be useful when the system is treated numerically,
assuming a finite size chain. It is well known that quantum spin chains
are typically investigated using a finite size system. However, here
we warn that this analysis must be treated carefully because for $N$
large enough, expressions \eqref{eq:C_sclng} and \eqref{eq:X_sclng}
fail, so we must use the more general relations \eqref{eq:CN-ex}
and \eqref{eq:CN-tan}, respectively.

\section{Ising-Heisenberg tetrahedral chain}

Quantum manifestations provided for instance by several real magnetic
materials, which can be viewed as one-dimensional systems. Like 3D
compounds in which, when we consider one columnar stripe, we could
observe a double tetrahedral chain structure. Such as cobalt oxide
$\mathrm{RBaCo_{4}}\mathrm{O}_{7}$, where $\mathrm{R}$ denotes a
rare earth atom, which has a swedenborgite lattice structure\citep{fritz}.
Another compound with a similar structure could be the salt with 3D
corrugated packing frustrated spin\citep{otsuka} of $\mathrm{C_{60}^{\bullet-}}$
in ($\mathrm{MDABCO^{+})(C_{60}^{\bullet-})}$ {[}$\mathrm{MDABCO^{+}}=N$-methyldiazabicyclooctanium{]}
cation and $\mathrm{C_{60}^{\bullet-}}$ radical anions, a stripe
of this salt can be viewed also as a double-tetrahedral chain.

\begin{figure}[h]
\centering{}\includegraphics[scale=0.8]{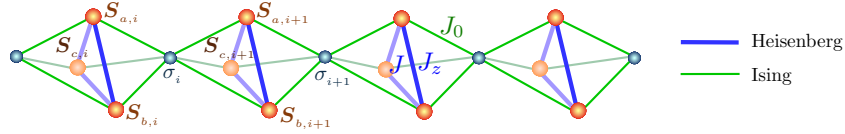}\caption{\label{fig:Sch-trd-ch}Schematic representation of Ising-Heisenberg
tetrahedral chain. Small balls ($\sigma_{i}$) corresponds to Ising
spins, and large balls ($\boldsymbol{S}_{a(b),i}$) correspond to
Heisenberg spins.}
\end{figure}

Initially, the Heisenberg tetrahedral chain was studied in reference
\citep{mambrini,roj-alc}. Later, the Ising-Heisenberg version of
the coupled tetrahedral chain was investigated earlier in reference
\citep{vadim-1,Vadim-2}. However, here we discuss a slightly different
Ising-Heisenberg tetrahedral chain (as illustrated in fig.\ref{fig:Sch-trd-ch}),
the thermodynamic properties in the neighbors of the pseudo-transition,
and for a finite size chain. Thus, the Hamiltonian of the model can
be expressed as
\begin{alignat}{1}
H= & -\sum_{i=1}^{N}\left\{ J(\boldsymbol{S}_{a,i},\boldsymbol{S}_{b,i})_{z}+J(\boldsymbol{S}_{b,i},\boldsymbol{S}_{c,i})_{z}+J(\boldsymbol{S}_{c,i},\boldsymbol{S}_{a,i})_{z}+\tfrac{h}{2}\left(\sigma_{i}+\sigma_{i+1}\right)\right.\nonumber \\
 & \left.\hspace{1cm}+\left(S_{a,i}^{z}+S_{b,i}^{z}+S_{c,i}^{z}\right)\left[h_{z}+J_{0}(\sigma_{i}+\sigma_{i+1})\right]\right\} ,\label{eq:H-tetra}
\end{alignat}
where $J(\boldsymbol{S}_{a,i},\boldsymbol{S}_{b,i})_{z}\equiv JS_{a,i}^{x}S_{b,i}^{x}+JS_{a,i}^{y}S_{b,i}^{y}+J_{z}S_{a,i}^{z}S_{b,i}^{z}$.
With $S_{a,i}^{\alpha}$ denoting Heisenberg spin-1/2, and $\alpha=\{x,y,z\}$,
while $\sigma_{i}$ denotes the Ising spin ($\sigma_{i}=\pm\frac{1}{2}$).
Similarly we define for sites $b$ and $c$ in \eqref{eq:H-tetra}.

\subsection{Zero temperature phase transition}

Below, we present for didactic reading only a revisit of the zero-temperature
phase transition. This model exhibits a peculiar phase transition
at zero temperature\citep{ph-bd}, in which the finite temperature
in the vicinity of this phase transition becomes a pseudo-transition.
We focus on the peculiar state of zero temperature, the ferrimagnetic
(FI) and frustrated ($FR_{2}$) state. Thus, the ferrimagnetic (FI)
states can be expressed as 
\begin{equation}
|FI\rangle=\prod_{i=1}^{N}\left|\substack{+\\
+\\
+
}
\right\rangle _{i}|-\rangle_{i},\label{Trt-FI}
\end{equation}
with Ising spin magnetization $m_{I}=-\frac{1}{2}$, Heisenberg spin
magnetization $m_{H}=\frac{1}{2}$ and total magnetization $m_{t}=1$.
The corresponding energy in the ground state of the ferrimagnetic
phase becomes 
\begin{alignat}{1}
E_{FI}= & \frac{1}{2}\left(3J_{0}+h\right)-\frac{3J_{z}}{4}-\frac{3h_{z}}{2}.
\end{alignat}
It is certainly worth mentioning that the $FI$ state has zero residual
entropy ($\mathcal{S}=0$) at zero temperature.

The other state that we are interested in is a frustrated ($FR_{2}$)
phase, expressed by 
\begin{equation}
|FR_{2}\rangle=\prod_{i=1}^{N}\left|\tfrac{1}{2},+\tfrac{1}{2}\right\rangle _{i}|+\rangle_{i},\label{Trt-FR2}
\end{equation}
where 
\begin{alignat*}{1}
\left|\tfrac{1}{2},+\tfrac{1}{2}\right\rangle _{i}= & \tfrac{1}{\sqrt{6}}\left(\left|\substack{+\\
+\\
-
}
\right\rangle _{i}-2\left|\substack{+\\
-\\
+
}
\right\rangle _{i}+\left|\substack{-\\
+\\
+
}
\right\rangle _{i}\right)\quad\text{or}\quad\tfrac{1}{\sqrt{2}}\left(\left|\substack{-\\
+\\
+
}
\right\rangle _{i}-\left|\substack{+\\
+\\
-
}
\right\rangle _{i}\right),
\end{alignat*}
so the state $\left|\tfrac{1}{2},+\tfrac{1}{2}\right\rangle _{i}$
is the responsible for the rise of frustration. And the corresponding
magnetizations are $m_{I}=\frac{1}{2}$, $m_{H}=\frac{1}{6}$ and
$m_{t}=1$. Thereby, its respective frustrated ground-state energy
becomes 
\begin{alignat}{1}
E_{_{FR_{2}}}= & -\frac{1}{2}\left(J_{0}+h\right)+\frac{J}{2}+\frac{J_{z}}{4}-\frac{h_{z}}{2}.
\end{alignat}
Whereas $FR_{2}$ state has residual entropy $\mathcal{S}=\ln(2)$
at a zero temperature in units of the Boltzmann constant $k_{B}$.

\subsection{Thermodynamics of Ising-Heisenberg tetrahedral chain}

In order to study the thermodynamic properties of the Hamiltonian
\eqref{eq:H-tetra}, we can solve this model through the transfer
matrix technique. Hence the transfer matrix has the following form
$\mathbf{V}=\left[\begin{array}{cc}
v_{1,1} & v_{1,2}\\
v_{1,2} & v_{2,2}
\end{array}\right]$, whose transfer matrix elements are denoted as $w_{1}=v_{1,1}$,
$w_{-1}=v_{2,2}$ and $w_{0}=v_{1,2}$, which are explicitly expressed
by
\begin{alignat}{1}
w_{n}= & 2{\rm e}^{\beta\left(\frac{2nh-J_{z}}{4}\right)}\left\{ \left({\rm e}^{\beta J}+2{\rm e}^{-\beta\frac{J}{2}}\right)\cosh\left(\beta\tfrac{nJ_{0}+h_{z}}{2}\right)+{\rm e}^{\beta J_{z}}\cosh\left(3\beta(\tfrac{nJ_{0}+h_{z}}{2})\right)\right\} 
\end{alignat}
where $n=\{-1,0,1\}$. With being $\beta=1/k_{B}T$, while $k_{B}$
denotes the Boltzmann constant, and $T$ is the absolute temperature.

Afterward, the eigenvalues of the transfer matrix become
\begin{equation}
\lambda_{1,2}=\tfrac{1}{2}\Bigl(w_{1}+w_{-1}\pm\sqrt{(w_{1}-w_{-1})^{2}+4w_{0}^{2}}\Bigr).\label{eq:eigvls}
\end{equation}
With this result in our hands, we can express the partition function
for the finite size chain as follows
\begin{equation}
\mathcal{Z}_{N}=\lambda_{1}^{N}+\lambda_{2}^{N}.
\end{equation}
Using the free energy per unit cell presented in \eqref{eq:free-f-N}
results in 
\begin{equation}
f^{(N)}=-\tfrac{1}{\beta}\ln(\lambda_{1})-\tfrac{1}{N\beta}\ln\left[1+\left(\frac{\lambda_{2}}{\lambda_{1}}\right)^{N}\right].\label{eq:free-energ-N}
\end{equation}
Whereas, the free energy \eqref{eq:f_e} in the thermodynamic limit
($N\rightarrow\infty$) reduces to 
\begin{equation}
f=-\tfrac{1}{\beta}\ln\left[\tfrac{1}{2}\Bigl(w_{1}+w_{-1}+\sqrt{(w_{1}-w_{-1})^{2}+4w_{0}^{2}}\Bigr)\right].\label{eq:free-energ}
\end{equation}
This result may indicate the presence of a genuine phase transition
at finite temperature when $w_{0}=0$, which also means degenerate
eigenvalues, since, at this limit, we have $\lambda_{2}=\lambda_{1}$.

The phase boundary ($E_{FI}=E_{FR_{2}}$) for $h_{z}=h$, restrict
the parameters to $J_{z}=2J_{0}-J/2$.

\subsection{Finite size effects on pseudo-transition}

It is worth mentioning that the residual entropy of the unusual phase
boundary at the interface of $FI$ and $FR_{2}$ is given by ${\cal S}=\ln(2)$.
The finite temperature pseudo-transition occurs as a consequence of
the zero-temperature phase transition between the ferrimagnetic ($FI$)
and frustrated ($FR_{2}$) phase\citep{ph-bd}.

Here we discuss the pseudo-transition of Ising-Heisenberg tetrahedral
chain from a different perspective. For this purpose, we write the
free energy in the vicinity of pseudo-critical temperature $T_{p}$.
As we know from the previous result investigated in reference \citep{pseudo},
the pseudo-critical temperature must be obtained using the following
relation 
\begin{equation}
w_{1}(T_{p})=w_{-1}(T_{p}),\label{eq:w1-w2}
\end{equation}
where $T_{p}$ corresponds to the pseudo-critical temperature. However, it
is evident that the condition of $\lambda_{1}>\lambda_{2}$ is always
satisfied, which guarantees the analyticity of free energy, although
pseudo-transition should only occur when $\lambda_{2}\rightarrow\lambda_{1}$. 

\begin{figure}[h]
\includegraphics[scale=0.27]{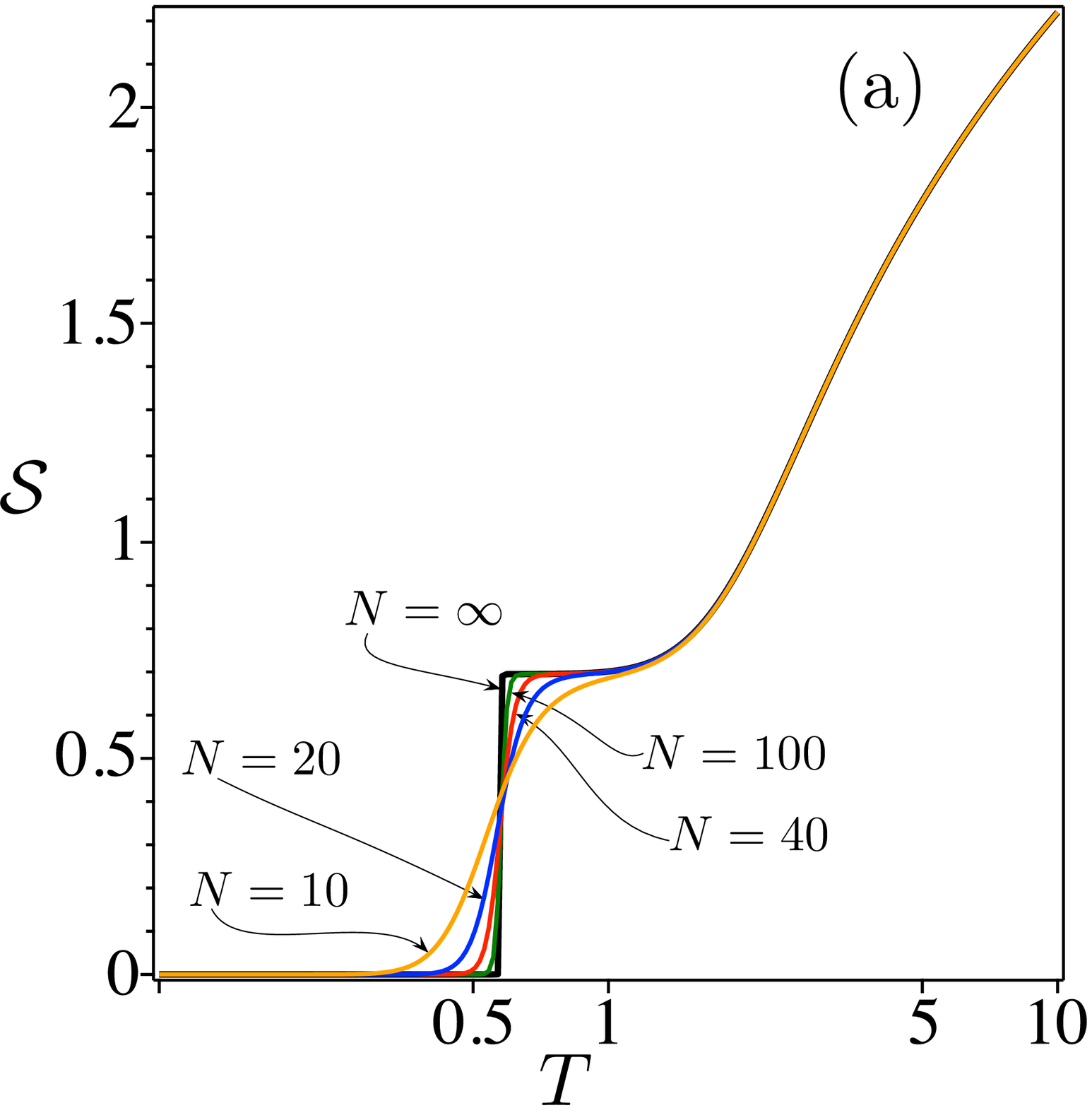}\includegraphics[scale=0.27]{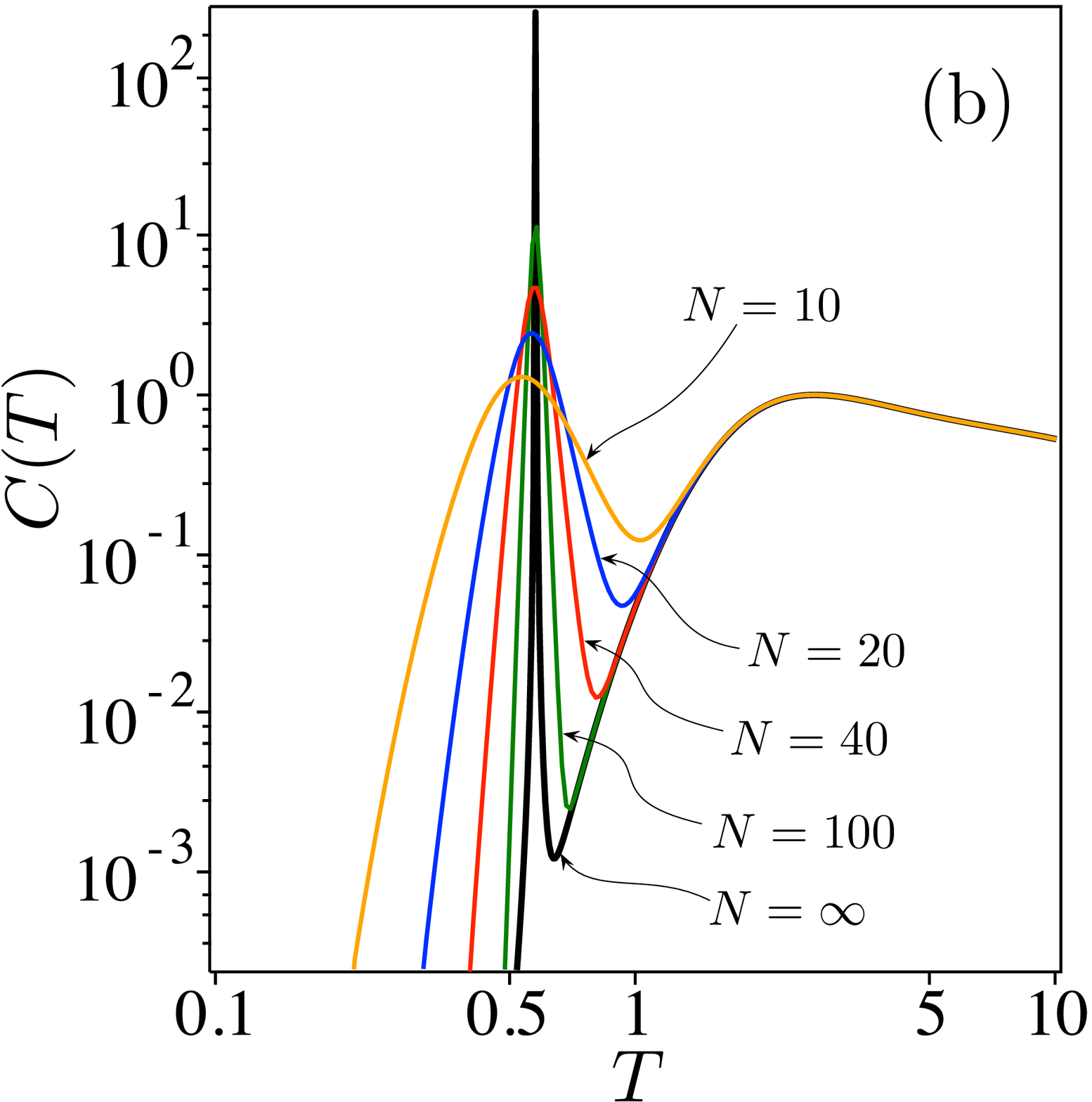}\includegraphics[scale=0.27]{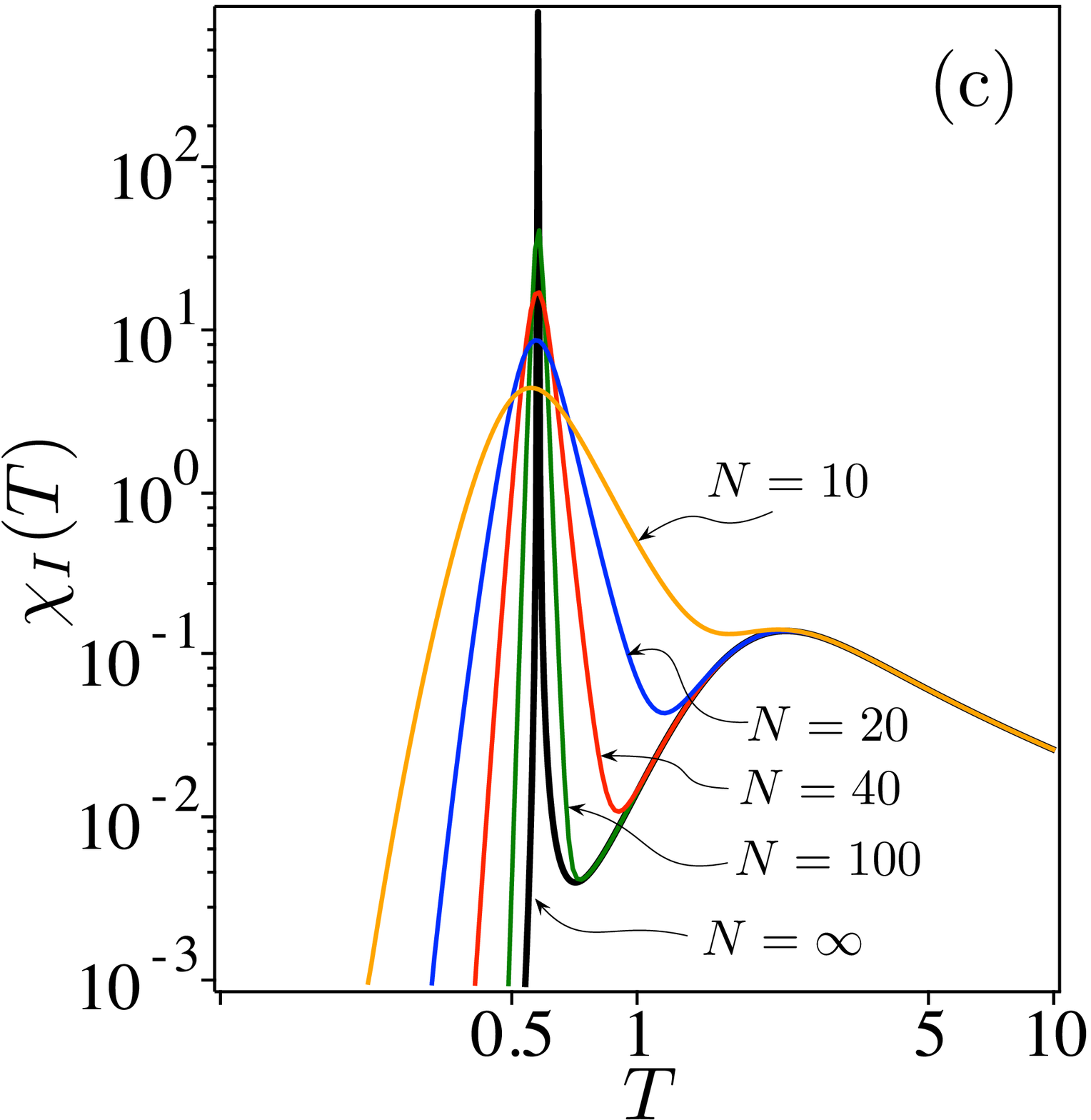}

\caption{\label{fig:(a)-Specific-heat}(a) Entropy as a function of temperature
for a set of finite size chains assuming fixed parameters $h=h_{z}=20$,
$J_{0}=-10$, $J=-10$ and $J_{z}=-14.6$. (b) Specific heat versus
$T$, for the same set of parameters in (a). In (c) is depicted the
Magnetic susceptibility as function of temperature, for the same set
of parameters in (a) and the same set of finite size chain. }
\end{figure}

In fig. \ref{fig:(a)-Specific-heat}a is illustrated the entropy $\mathcal{S}$
as a function of temperature for a set of values $N=\{10,20,40,100\}$
and parameters given in the figure legend. Where we observe a continuous
step function around pseudo-critical temperature $T_{p}$, for small
$N$ the corners of the step function are rounded, as far as $N$
increases, the rounded corners become increasingly sharp. In fig.\ref{fig:(a)-Specific-heat}b,
we report the specific heat $C(T)$ on a logarithmic scale as a function
of temperature, for the set of values in panel (a). It is clear how
the specific heat peak increases with the number of $C^{(N)}(T_{p})\propto N$
unit cells, and the location of peak shifts slightly, apparently indicating
a possible phase transition. The solid black line corresponds to an
infinite chain, exhibiting a sharp and robust peak that practically
appears to be divergent at a pseudo-critical temperature. The magnetic
susceptibility for the Ising spin ($\chi_{I}=-\frac{\partial^{2}f}{\partial h^{2}}$),
and the Heisenberg spin ($\chi_{H}=-\frac{\partial^{2}f}{\partial h_{z}^{2}}$)
are almost the same in the neighboring of the pseudo-critical temperature.
So we can just denote by $\chi(T_{p})\equiv\chi_{H}(T_{p})\thickapprox\chi_{I}(T_{p})$.
In fig.\ref{fig:(a)-Specific-heat}c, the magnetic susceptibility
on a logarithmic scale against temperature is reported. In principle,
we have a property similar to that illustrated in panel (b).

\begin{figure}[h]
\includegraphics[scale=0.25]{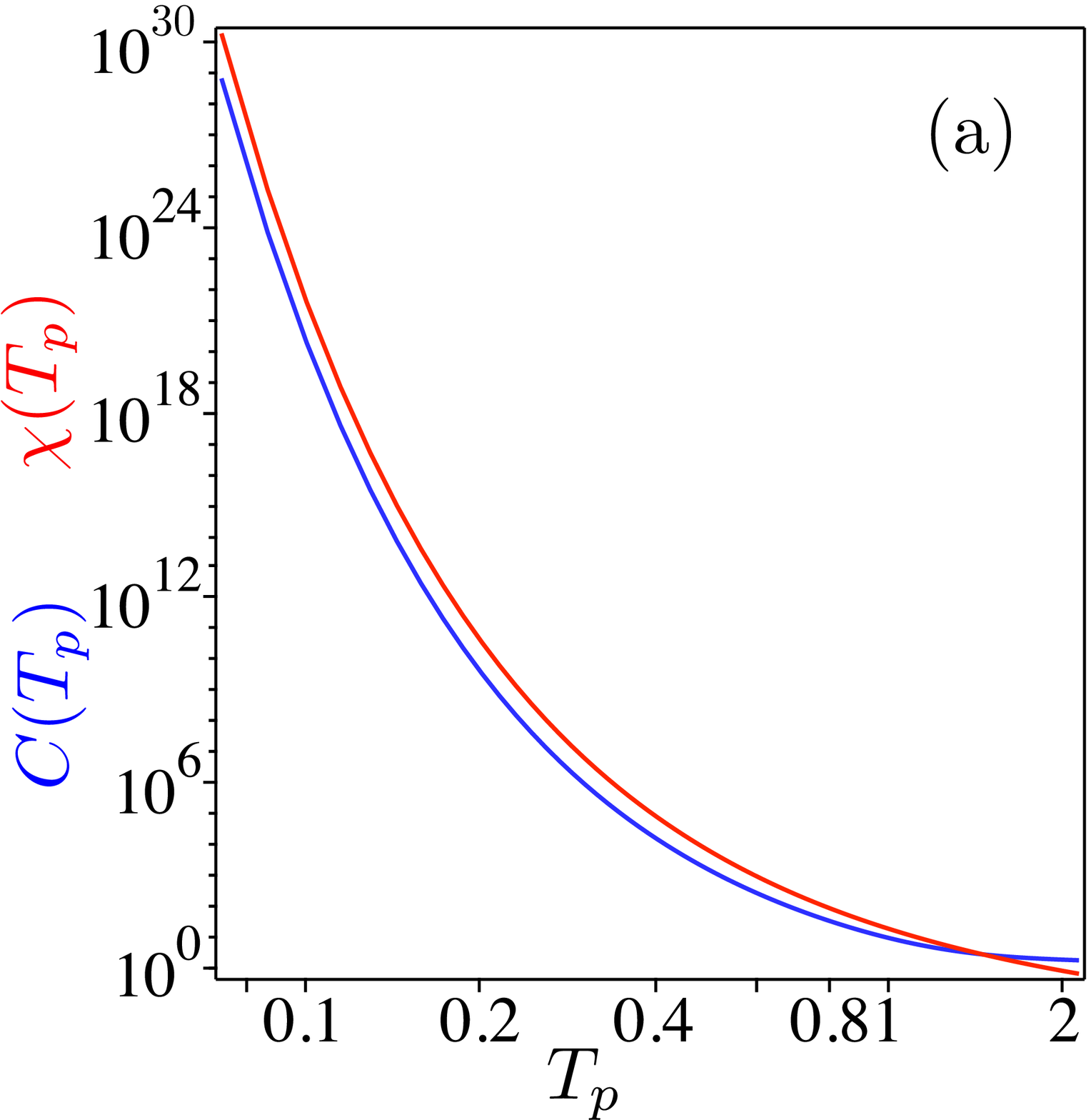} \enskip{}\includegraphics[scale=0.25]{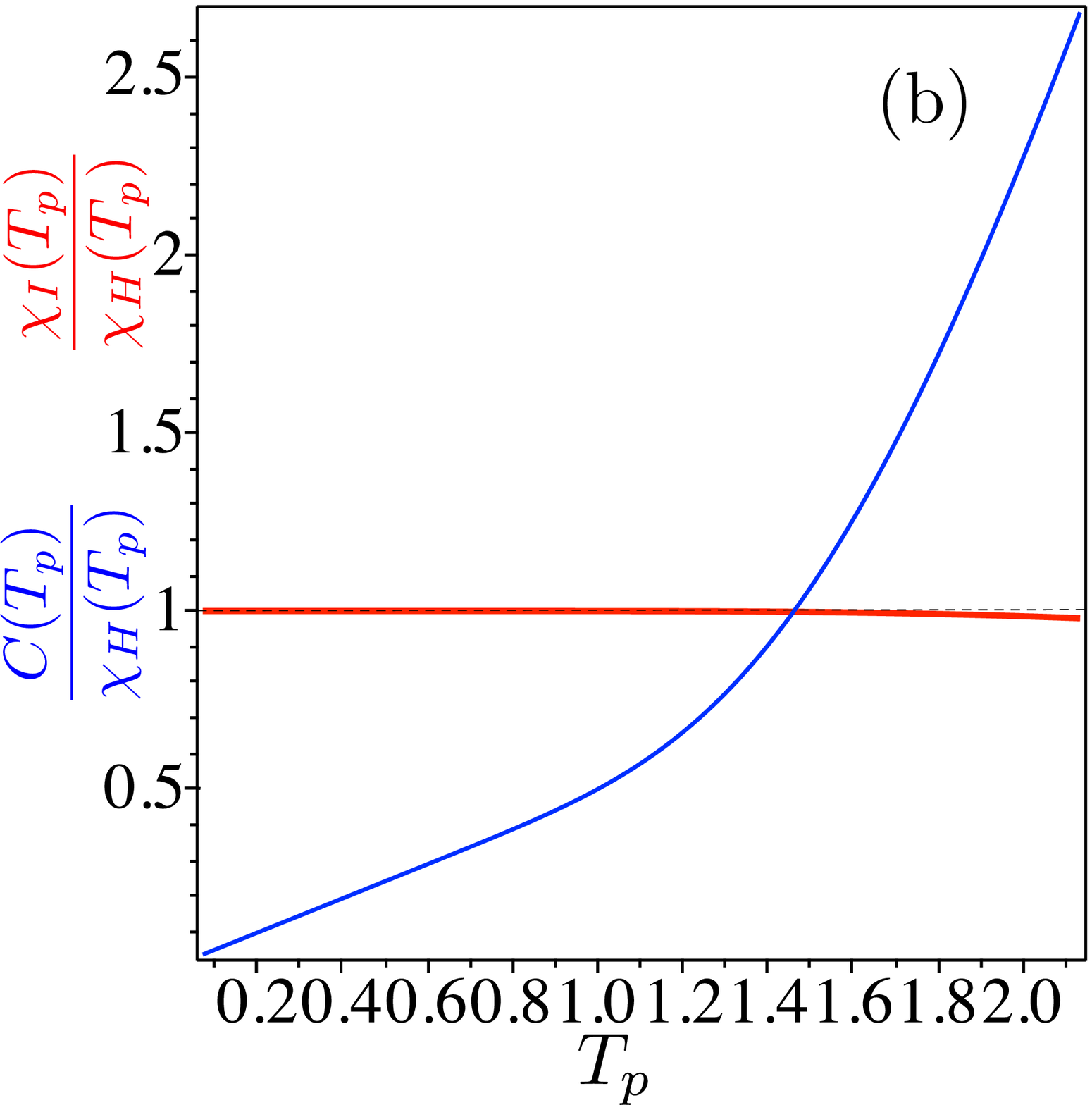}\caption{\label{fig:peaks}(a) $C(T_{p})$ and $\chi(T_{p})$ as a function
of $T_{p}$ in logarithmic scale. (b) $C(T_{p})/\chi_{H}(T_{p})$
and $\chi_{I}(T_{p})/\chi_{H}(T_{p})$ as a function of temperature.
In both panels for fixed $h=h_{z}=20$, $J_{0}=-10$, $J=-10$ and
$J_{z}$ is restricted to $T_{p}$ by eq. \eqref{eq:w1-w2}.}
\end{figure}

In fig.\ref{fig:peaks}a, the $C(T_{p})$ is depicted as a function
of $T_{p}$, which shows how the height of specific heat increases
when $T_{p}$ decreases (blue line). However, when $T_{p}\rightarrow0$,
we have $C(T_{p})\rightarrow\infty$ then we must have a real phase
transition only at $T_{p}=0$. Furthermore, magnetic susceptibility
$\chi(T_{p})$ is illustrated as dependence on $T_{p}$, which increases
monotonically when the pseudo-critical temperature decreases. The
peaks are really huge in the low-temperature region (as $T_{p}\neq0$),
but it is still only a finite peak.

To illustrate an additional property of $C(T_{p})$, $\chi(T_{p})$
and $\chi_{H}(T_{p})$ we report in fig.\ref{fig:peaks}b the ration
of $C(T_{p})/\chi_{H}(T_{p})$ as a function of $T_{p}$ represented
by a solid blue line, which indicates both quantities are of the same
order. Analogously the ratio $\chi_{I}(T_{p})/\chi_{H}(T_{p})$ is
depicted by a red line, which reveals almost a constant value around
1; this confirms that both quantities are quite similar around the
pseudo-critical peak.

Indeed, this result is valid only in the neighboring of pseudo-critical
temperature.

\begin{figure}[h]
\includegraphics[scale=0.25]{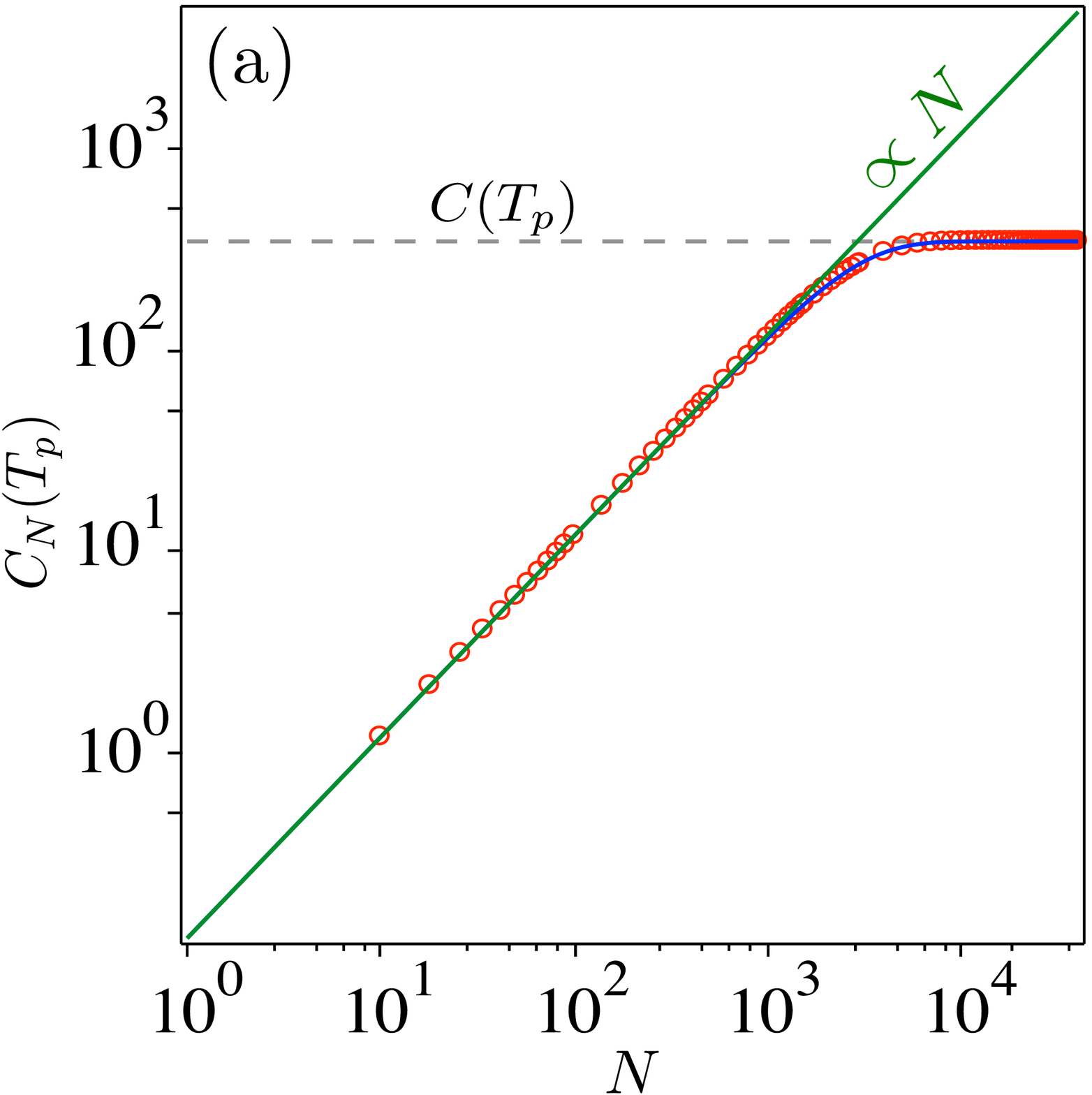} \includegraphics[scale=0.25]{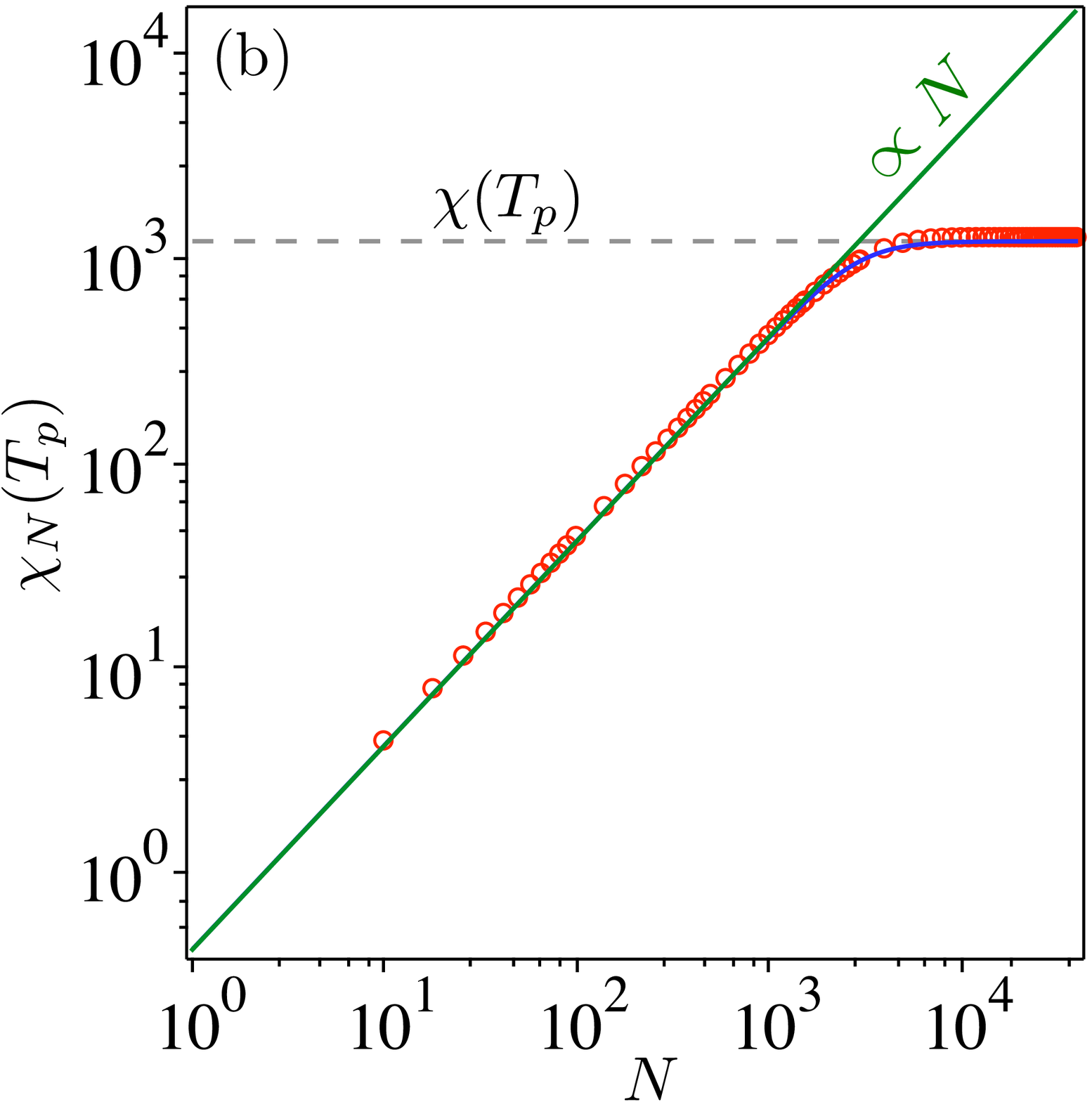}\caption{\label{fig:numeric-CX}(a) Specific heat peak in logarithmic scale
$C^{(N)}(T_{p})$ as a function of $N$ described by circled data
(shows only few values), blue solid line corresponds to non-linear
fitting curve \eqref{eq:CNp}, and green solid line corresponds to
$C^{(N)}(T_{p})\propto N$. (b) Magnetic susceptibility $\chi_{N}(T_{p})$
in logarithmic scale as a function of $N$ described by circled data
(shows only few values), blue solid line represents the curve \eqref{eq:XNp},
and solid green line denotes $\chi_{N}(T_{p})\propto N$. Assuming
in both panels fixed parameters $h=h_{z}=20$, $J_{0}=-10$, $J=-10$
and $J_{z}=-14.6$.}
\end{figure}

Last but not least, in fig.\ref{fig:numeric-CX}a, we illustrate the
specific heat for a finite chain $C^{(N)}(T_{p})$ as a function of
$N$ in a logarithmic scale, this curve is denoted by circled data
obtained from the exact result. The blue line curve describes the
function \eqref{eq:CN-ex}, which satisfactorily fits with an exact
result for all values of $N$. While green line stands for the straight
line given by the relation \eqref{eq:C_sclng}, and we observe that
for $N\lesssim10^{3}$ the specific heat increases proportionally
to $N$, but for $N\gtrsim10^{3}$ fails. 

Alternatively, we present a nonlinear empirical function that can
also be considered to fit nicely with the exact curve, which is given
below 
\begin{equation}
C^{(N)}(T_{p})\thickapprox\left(1+\frac{a}{N^{3}}\right)^{-\frac{1}{3}}C(T_{p}),\label{eq:CNp}
\end{equation}
where $a\thickapprox2.829\times10^{10}$ is a fitted constant for
a fixed $T_{p}=0.577077991$. For a moderately large $N\lesssim10^{3}$,
the empirical expression reduces to $C^{(N)}(T_{p})\thicksim N\,C(T_{p})/\sqrt[3]{a}$,
just observing at moderately large $N$, a divergence could be induced
when $N\rightarrow\infty$, meaning a possible singularity at $T_{p}$.
However, eqs.\eqref{eq:CN-ex} and \eqref{eq:CNp} give us the peak
dependence for all range of $N$, which at $N\rightarrow\infty$,
just leads to a finite peak instead of a divergence, this shows that
there is no genuine phase transition.

Similarly, in fig.\ref{fig:numeric-CX}b, we illustrate the magnetic
susceptibility $\chi_{N}(T_{p})$ as a function of $N$ denoted by
circles. The curves are fairly similar to the panel (a), then an empirical
fit curve also gives us
\begin{equation}
\chi_{N}(T_{p})\thickapprox\left(1+\frac{a}{N^{3}}\right)^{-\frac{1}{3}}\chi(T_{p}),\label{eq:XNp}
\end{equation}
with $a\thickapprox2.815\times10^{10}$ being a fitted constant for
a given $T_{p}=0.577077991$. Again, for moderately large $N\apprle10^{3}$
the eq.\eqref{eq:XNp} reduces to $\chi_{N}(T_{p})\thicksim N\,\chi(T_{p})/\sqrt[3]{a}$
which fits nicely, although for $N\apprge10^{3}$ considerably large,
it fails.

\section{Ising-Heisenberg ladder model}

Other systems of interest are the spin-1/2 quantum Heisenberg ladder\citep{Betchelor07}.
Some of the most widespread compounds in the spin-1/2 Heisenberg ladder
materials literature are cuprates $\mathrm{Cu_{2}(C_{5}H_{12}N)_{2}Cl_{4}}$\citep{Chiari},
$\mathrm{SrCu_{2}O_{3}}$\citep{Hiroi}, $\mathrm{(C_{5}H_{12}N)_{2}CuBr_{4}}$\citep{Willett},
and vanadates $\mathrm{M^{2+}V_{2}O_{5}}$ \citep{Onoda}, $\mathrm{(VO)_{2}P_{2}O_{7}}$
\citep{barnes}, which involve $\mathrm{Cu^{2+}}$ and $\mathrm{V^{4+}}$
magnetic ions represented as the spin-1/2 particles. Another compound
that is well described by spin-1/2 Heisenberg two-leg ladder is $\mathrm{Cu(Qnx)(Cl_{1-\mathit{x}}Br_{\mathit{x}})_{2}}$
, where $\mathrm{Qnx}$ stands for quinoxaline ($\mathrm{C_{8}H_{6}N_{2}}$)\citep{Simutis}.

As a second application, we consider the model studied in reference
\citep{on-strk}, although this model has been considered previously,
the properties discussed here have not been studied before.

\begin{figure}[h]
\includegraphics[scale=0.8]{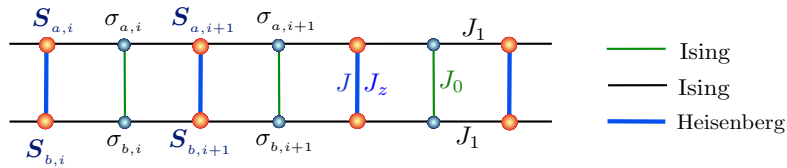}

\caption{\label{fig:ladder} A schematic representation of the spin-1/2 Ising-Heisenberg
ladder model with alternating Ising and Heisenberg inter-leg interactions,
thick vertical lines correspond to the Heisenberg coupling ($J,J_{z}$),
whereas thin vertical and horizontal lines correspond to the Ising
exchange interactions $J_{0}$ and $J_{1}$.}

\end{figure}

Therefore, let us consider the spin-1/2 Ising-Heisenberg ladder with
alternating Ising and Heisenberg inter-leg couplings and the Ising
intra-leg coupling schematically depicted in figure \ref{fig:ladder}.
The Hamiltonian described above for the spin-1/2 Ising-Heisenberg
ladder is given by 
\begin{equation}
\mathcal{H}=-\sum_{i=1}^{N}\left[J(\boldsymbol{S}_{a,i},\boldsymbol{S}_{b,i})_{z}+\frac{J_{0}}{2}(\sigma_{a,i}\sigma_{b,i}+\sigma_{a,i+1}\sigma_{b,i+1})+\sum_{\gamma=a,b}J_{1}(\sigma_{\gamma,i}+\sigma_{\gamma,i+1})S_{\gamma,i}^{z}\right],\label{eq:Ham-orig}
\end{equation}
where $J(\boldsymbol{S}_{a,i},\boldsymbol{S}_{b,i})_{z}=J(S_{a,i}^{x}S_{b,i}^{x}+S_{a,i}^{y}S_{b,i}^{y})+J_{z}S_{a,i}^{z}S_{b,i}^{z}$,
with $S_{\gamma,i}^{\alpha}$ denotes spatial components of the spin-1/2
operator ($\alpha=\{x,y,z\}$) at site $i$, and $\gamma=a$ or $b$
(see figure \ref{fig:ladder}). $J_{1}$ denotes the Ising inter-leg
coupling, similarly the Ising intra-leg coupling is denoted by $J_{0}$.
Whereas, the anisotropic XXZ Heisenberg inter-leg exchange interaction
has two spatial components $J$ and $J_{z}$ in the $xy$-plane and
along $z$-axis, respectively.

\subsection{Thermodynamics of Ising-Heisenberg ladder model}

To study the thermodynamics of the spin-1/2 Ising-Heisenberg ladder
with alternating inter-leg couplings, let us perform the partition
function given by \eqref{eq:part-func}.

The thermodynamic of the Ising-Heisenberg ladder model can be obtained
by the usual transfer matrix approach\citep{kramers}, which has the
form, 

\begin{equation}
\boldsymbol{V}=\left[\begin{array}{cccc}
v_{1,1} & v_{1,2} & v_{1,2} & v_{1,4}\\
v_{1,2} & v_{2,2} & v_{1,4}u^{-2} & v_{1,2}\\
v_{1,2} & v_{1,4}u^{-2} & w_{2,2} & v_{1,2}\\
v_{1,4} & v_{1,2} & v_{1,2} & v_{1,1}
\end{array}\right],\label{eq:T-trsnf}
\end{equation}
whose transfer matrix elements are given explicitly by 
\begin{alignat}{1}
v_{1,1}= & uz\left(y^{4}+y^{-4}\right)+\frac{u}{z}\left(x^{2}+x^{-2}\right),\\
v_{1,2}= & z\left(y^{2}+y^{-2}\right)+z^{-1}(y_{1}^{2}+y_{1}^{-2}),\\
v_{1,4}= & 2uz+\frac{u}{z}\left(x^{2}+x^{-2}\right),\\
v_{2,2}= & 2\frac{z}{u}+\frac{1}{uz}\left(y_{2}^{2}+y_{2}^{-2}\right),
\end{alignat}
with $x=\mathrm{e}^{\beta J/4}$ , $y=\mathrm{e}^{\beta J_{1}/4}$,
$z=\mathrm{e}^{\beta J_{z}/4}$, $u=\mathrm{e}^{\beta J_{0}/4}$,
and additionally, we define also the following exponential $y_{1}=\mathrm{e}^{\beta\sqrt{J^{2}+J_{0}^{2}}/4}$
and $y_{2}=\mathrm{e}^{\beta\sqrt{J^{2}+4J_{0}^{2}}/4}$.

To find the eigenvalues of the matrix \eqref{eq:T-trsnf}, we proceed
to calculate the $\text{det}(\boldsymbol{V}-\lambda)=0$. Then, the
determinant falls into a secular fourth order equation in $\lambda$.
Factoring this polynomial, we obtain the following expression 
\begin{alignat}{1}
0= & \left(\lambda-v_{2,2}+v_{1,4}u^{-2}\right)\left(\lambda-v_{1,1}+v_{1,4}\right)\times\nonumber \\
 & \left[\lambda^{2}-\left(v_{1,1}+v_{1,4}+v_{2,2}+v_{1,4}u^{-2}\right)\lambda+\left(v_{1,1}+v_{1,4}\right)\left(v_{2,2}+v_{1,4}u^{-2}\right)-\left(2v_{1,2}\right)^{2}\right].\label{eq:cubic-eq}
\end{alignat}

After that, the corresponding eigenvalues are expressed as follows
\begin{align}
\lambda_{1}= & \frac{1}{2}\left(w_{1}+w_{-1}+\sqrt{\left(w_{1}-w_{-1}\right)^{2}+4w_{0}^{2}}\right),\label{eq:L-lmb1}\\
\lambda_{2}= & \frac{1}{2}\left(w_{1}+w_{-1}-\sqrt{\left(w_{1}-w_{-1}\right)^{2}+4w_{0}^{2}}\right),\\
\lambda_{3}= & v_{2,2}-v_{1,4}u^{-2},\\
\lambda_{4}= & v_{1,1}-v_{1,4},
\end{align}
where

\begin{alignat}{1}
w_{1}= & v_{1,1}+v_{1,4},\label{eq:Lw1}\\
w_{-1}= & v_{2,2}+v_{1,4}u^{-2},\\
w_{0}= & 2v_{1,2}.\label{eq:Lw0}
\end{alignat}
In general, as previously assumed, the elements satisfy $v_{i,j}>0$.
It can be easily seen that the first eigenvalue (\ref{eq:L-lmb1})
is always a positive amount and represents the largest eigenvalue
of the transfer matrix. It is obvious to verify that $\lambda_{1}>\lambda_{2}$,
then $\lambda_{1}$ is the largest eigenvalue, using the same procedure
to that found in ref.\citep{ky lin,Ninio}, it is evident that $\lambda_{2}$
is the second largest eigenvalue.

Furthermore, we can write the relations (\ref{eq:Lw1}-\ref{eq:Lw0})
explicitly in terms of Hamiltonian parameter
\begin{align}
w_{1}= & 2{\rm e}^{\beta\frac{J_{0}+J_{z}}{4}}{\rm ch}\left(\tfrac{\beta J_{1}}{2}\right)^{2}+2{\rm e}^{\beta\frac{J_{0}-J_{z}}{4}}{\rm ch}\left(\tfrac{\beta J}{2}\right),\label{eq:cf-A}\\
w_{-1}= & {\rm e}^{\frac{-\beta\left(J_{0}+J_{z}\right)}{4}}\left[{\rm ch}\left(\tfrac{\beta}{2}\sqrt{J^{2}+4J_{1}^{2}}\right)+{\rm ch}\left(\tfrac{\beta J}{2}\right)\right]+2{\rm e}^{\frac{-\beta\left(J_{0}-J_{z}\right)}{4}},\\
w_{0}= & 2{\rm e}^{\beta\frac{J_{z}}{4}}{\rm ch}\left(\tfrac{\beta J_{1}}{2}\right)+2{\rm e}^{-\beta\frac{J_{z}}{4}}{\rm ch}\left(\tfrac{\beta}{2}\sqrt{J^{2}+J_{1}^{2}}\right).\label{eq:cf-C}
\end{align}

In the thermodynamic limit $N\to\infty$, the free energy per unit
cell \eqref{eq:f_e} is given only by the largest transfer-matrix
eigenvalue, where $w_{1}$, $w_{-1}$, and $w_{0}$ are given by eqs.(\ref{eq:cf-A})-(\ref{eq:cf-C}).
Quantities like entropy or specific heat can be obtained merely from
the free energy \eqref{eq:f_e}, by using the standard thermodynamic
formulas.

\subsection{Finite size effects on pseudo-transition}

The Ising ladder model with Ising-Heisenberg intra-rung coupling is
characterized indeed by the Hamiltonian \eqref{eq:Ham-orig}, here
we explore the phase boundary between two particular phases. These
states can be expressed as follows: 

The first one is the frustrated phase, which is denoted by 
\begin{align}
|FRU_{1}\rangle= & \prod_{i=1}^{N}|\tau_{0}\rangle_{i}\otimes|\begin{smallmatrix}\sigma_{1}\\
\sigma_{1}
\end{smallmatrix}\rangle_{i},\label{eq:SS}
\end{align}
with $|\tau_{0}\rangle_{i}=\frac{1}{\sqrt{2}}\left(|\begin{smallmatrix}+\\
-
\end{smallmatrix}\rangle_{i}+|\begin{smallmatrix}-\\
+
\end{smallmatrix}\rangle_{i}\right)$ and $\sigma_{1}$ denotes Ising spin orientation, which yields a
macroscopically degenerate state. Whereas the corresponding eigenvalue
is given by
\begin{align}
E_{FRU_{1}}= & -\frac{1}{2}|J|+\frac{1}{4}J_{z}-\frac{1}{4}J_{0}.
\end{align}
The above frustrated phase has a residual entropy $\mathcal{S}=\ln(2)$.

The other state is the antiferromagnetic (AFM) state, which is represented
by 
\begin{align}
|AFM\rangle= & \prod_{i=1}^{N}|\eta\rangle_{i}\otimes|\begin{smallmatrix}+\\
-
\end{smallmatrix}\rangle_{i},
\end{align}
with 
\begin{equation}
|\eta\rangle_{i}=\frac{\left(|\begin{smallmatrix}+\\
-
\end{smallmatrix}\rangle_{i}+c|\begin{smallmatrix}-\\
+
\end{smallmatrix}\rangle_{i}\right)}{\sqrt{1+c^{2}}},
\end{equation}
and 
\begin{equation}
c=\frac{\sqrt{4J_{1}^{2}+J^{2}}+2J_{1}}{J}.
\end{equation}

The corresponding antiferromagnetic eigenvalue results in
\begin{align}
E_{AFM}= & -\frac{1}{4}J_{z}-\frac{1}{4}J_{0}-\frac{1}{2}\sqrt{4J_{1}^{2}+J^{2}}.
\end{align}
Obviously, there is no residual entropy for the antiferromagnetic
phase.

The phase boundary between $FRU_{1}$ and $AFM$ has a peculiar interface,
since the residual entropy at the border is $\mathcal{S}=\ln(2)$,
which becomes the critical residual entropy of the frustrated phase
$\mathcal{S}=\ln(2)$, making the residual entropy a continuous function
at the phase boundary\citep{ph-bd}. That is, at this limit, we must
observe pseudo-transition at finite temperature. 

When the eigenvalues $\lambda_{1}$ and $\lambda_{2}$ becomes almost
degenerate ($\lambda_{2}\rightarrow\lambda_{1}$) in eq.\eqref{eq:eigvls},
then from the \eqref{eq:L-lmb1} we can get a pseudo-critical condition
when $w_{1}-w_{-1}=0$, in a similar way to that obtained in eq.\eqref{eq:w1-w2}.
Thus, this condition becomes as follows

\begin{equation}
v_{1,1}(T_{p})+v_{1,4}(T_{p})=v_{2,2}(T_{p})+v_{2,3}(T_{p}).
\end{equation}

\begin{figure}
\includegraphics[scale=0.27]{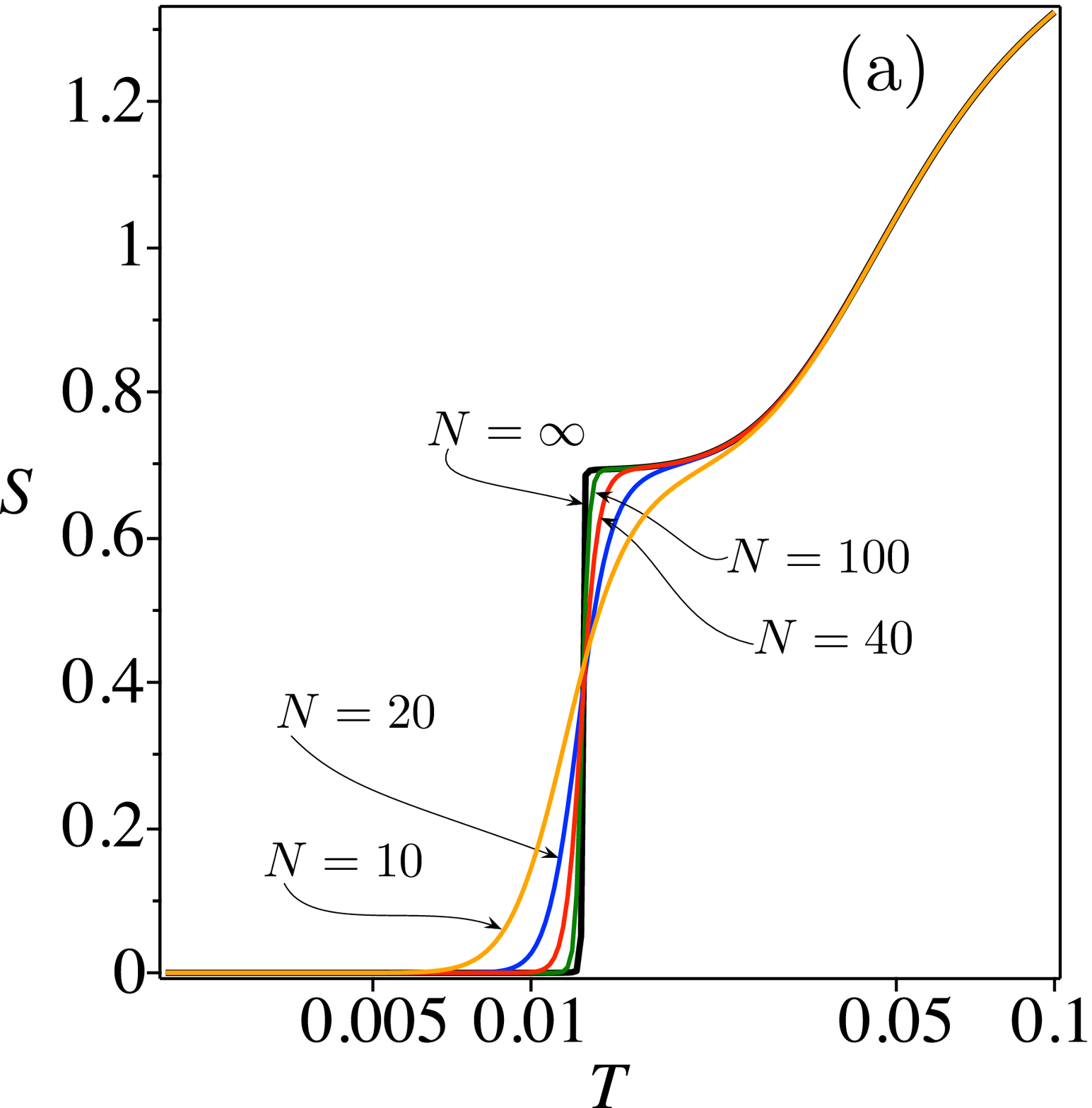}\includegraphics[scale=0.27]{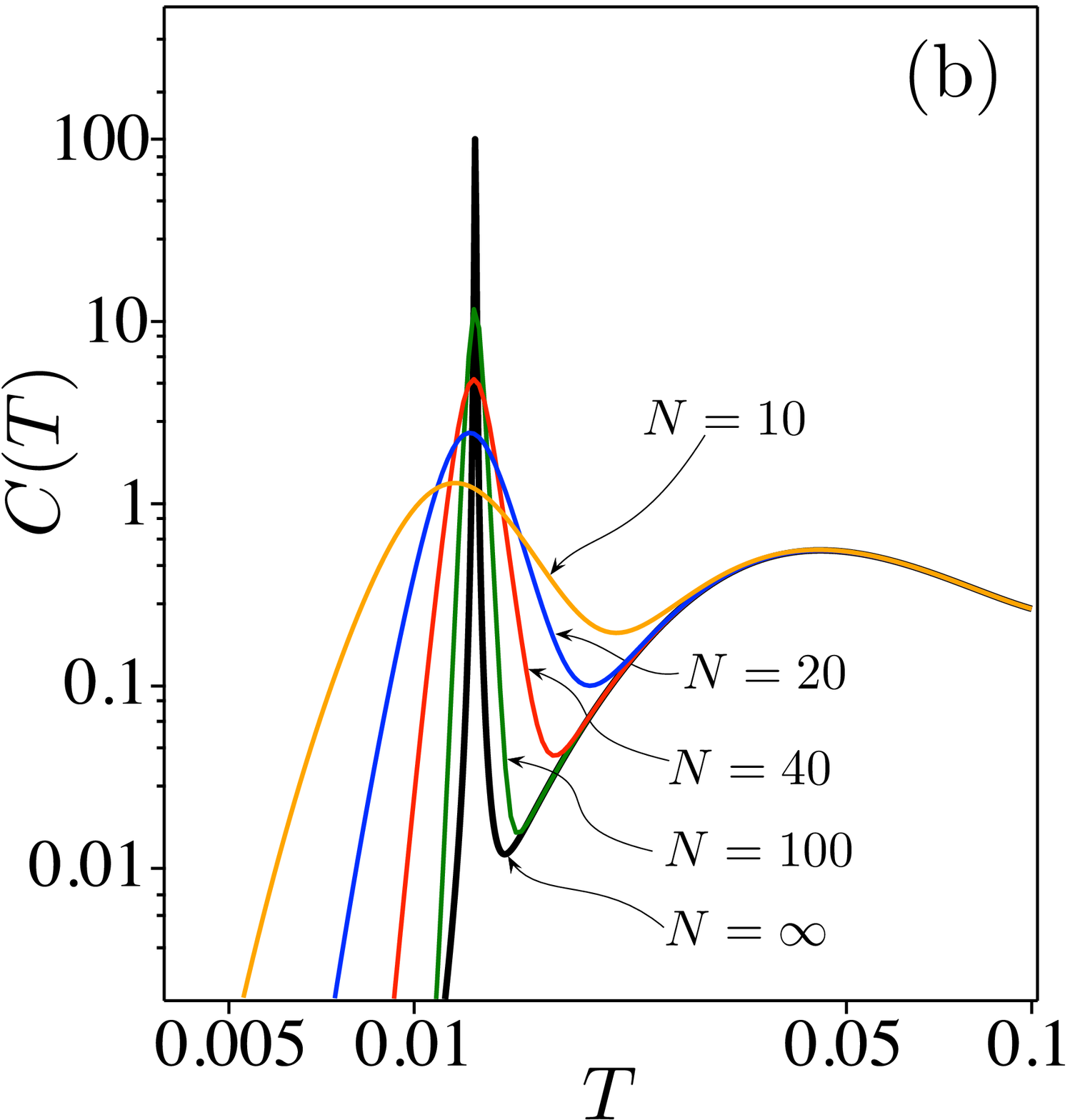}\includegraphics[scale=0.27]{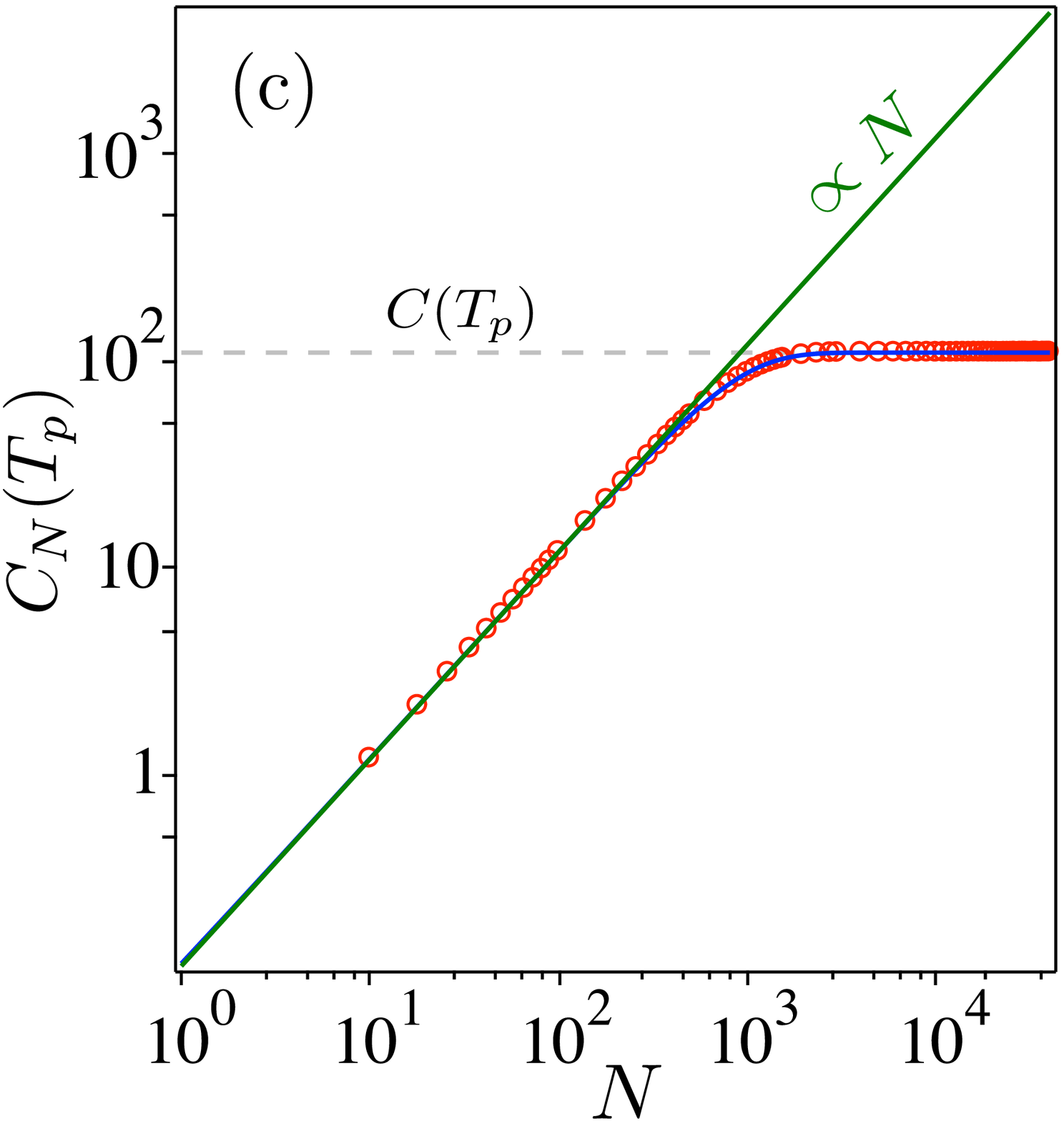}\caption{\label{fig:Specific-heat6} Entropy as a function of temperature,
for fixed parameters $J_{0}=1$, $J_{1}=\frac{\sqrt{3}}{2}+0.01$,
$J_{z}=-1$ and $J=1$; for several number of sites $N=\{10,20,40,100,\infty\}$.
(b) Specific heat as a function of temperature for the same set of
parameters considered in (a). In (c) the specific heat peak against
$N$ is reported in a logarithmic scale, for the same set of parameters
in (a), the green line illustrates $C^{(N)}\propto N$, blue line
describes the function \eqref{eq:CN-ex}, while circled curve stands
for the exact result. }

\end{figure}

In fig.\ref{fig:Specific-heat6}a is depicted the entropy as a function
of temperature for a finite size chains $N=\{10,20,40,100\}$ and
fixed parameters given in the legend. Where we observe a continuous
step function around pseudo-critical temperature $T_{p}$, as far
as $N$ increases, the rounded corners of step the function become
increasingly acute. In panel (b) is reported the specific heat as
a function of temperature for the same set of finite chains and parameters
given for (a). Once again, we can observe how the peak increases with
the number of sites $N$, around the pseudo-transition peak. The peak
height is clearly sensitive to $N$, rapidly converging to temperatures
higher than the pseudo-critical temperature. Moreover, for $N\rightarrow\infty$,
the peak becomes a very sharp peak, quite similar to a continuous
phase transition divergence. Whereas in fig.\ref{fig:Specific-heat6}c,
the depict the specific heat as a function of $N$, which was drawn
on logarithmic scale for convenience. The green line illustrates $C^{(N)}\propto N$
($C^{(N)}=\frac{C\zeta}{2}N$), blue line describes the function \eqref{eq:CN-ex},
while circled curve stands for the exact result. Indeed the specific
heat satisfies the finite size correction for $N\lesssim10^{3}$,
while for $N\gtrsim10^{3}$, the peak height leads to a constant value
$C(T_{p})$.

\section{Discussion and conclusion}

Few one-dimensional models can be solved exactly; in this case, it
is natural to study these models already in the thermodynamic limit.
But the vast majority of one-dimensional models have no exact solution,
so naturally numerical treatments are used to consider the finite
size lattice system. Hence, evidence of phase transitions in finite
size lattice systems appears as peaks in quantities that depend on
the second derivative of thermodynamic potentials. By performing extrapolation
techniques, we can identify that the peaks become divergences. Consequently,
there is a natural question: what is the relationship between the
height of the peak and the finite size chain? Although the one-dimensional
models we studied have already been examined in the thermodynamic
limit, like decorated models or ladder-type models, the second dimension
would manifest itself. In this sense, here we show that the pseudo
transitions discussed above have the same origin as finite lattice
models in a higher dimension than one, which justifies the name of
pseudo-transition. Therefore, we have applied to the Ising-Heisenberg
tetrahedral chain and the Ising-Heisenberg ladder model and investigated
its finite size effects around pseudo-critical temperature. For sure,
both models do not exhibit a genuine phase transition at finite temperature.
Despite this, an intense peak emerges in specific heat and magnetic
susceptibility at pseudo-critical temperature, closely resembling
a second-order phase transition. Still, the system does not exhibit
a genuine phase transition at finite temperature. For a moderately
large system, the peak depends proportionally on the number of the
unit cells. Although, for a sufficiently large system, the height
of the peak saturates at a finite value. For most one-dimensional
systems, the thermodynamic result is a cumbersome task to get exact
solutions. Since, it is common to simulate a finite size system computationally. 

The research to find the pseudo-critical temperature in quantum spin
chain models is a challenging topic, since, for most one-dimensional
models, we cannot get a precise thermodynamic function, which is necessary
to observe pseudo-transitions. There is a natural question: observe
the pseudo-critical temperature in pure quantum systems? In this sense,
our result and together with techniques proposed in reference \citep{ph-bd},
should be useful to shed light on non-trivial one-dimensional systems
such as quantum spin chain models, quantum ladder spin models, quantum
triangular tube spin models, one-dimensional Hubbard model\citep{Zhao,Zhang,Ding,Shi},
or other non-trivial quantum 1d-like systems.

\section*{Acknowledgment}

Work partially supported by Brazilian agency CNPq and FAPEMIG.


\begin{thebibliography}{99}
\bibitem{sarkanych}P. Sarkanych, Y. Holovatch and R. Kenna, Exact
solution of a classical short-range spin model with a phase transition
in one dimension: The Potts model with invisible states, Phys. Lett.
A 381 (2017), 3589.

\bibitem{kittel}C. Kittel, Phase Transition of a Molecular Zipper,
Am. J. Phys. 37 (1969) 917.

\bibitem{chui}S. T. Chui and J. D. Weeks, Pinning and roughening
of one-dimensional models of interfaces and steps, Phys. Rev. B 23
(1981) 2438.

\bibitem{dauxois}T. Dauxois and M. Peyrard, Entropy-driven transition
in a one-dimensional system, Phys. Rev. E 51 (1995) 4027.

\bibitem{Ninio} F. Ninio, A simple proof of the Perron-Frobenius
theorem for positive symmetric matrices, Phys. A: Math. Gen. 9 (1976)
1281.

\bibitem{Timonin}P. N. Timonin, Spin ice in a field: Quasi-phases
and pseudo-transitions, J. Exp. Theor. Phys. 113 (2011) 251.

\bibitem{Galisova}L. Galisova and J. Stre\v{c}ka,Vigorous thermal
excitations in a double-tetrahedral chain of localized Ising spins
and mobile electrons mimic a temperature-driven first-order phase
transition, Phys. Rev. E 91 (2015) 022134.

\bibitem{galisova17}L. Galisova, Magnetization plateau as a result
of the uniform and gradual electron doping in a coupled spin-electron
double-tetrahedral chain, Phys. Rev. E 96 (2017) 052110; Pairwise
Entanglement in Double-Tetrahedral Chain with Different Landé g-Factors
of the Ising and Heisenberg Spins, Acta Phys. Pol. A 137 (2020) 604.

\bibitem{strk-cav}J. Stre\v{c}ka, R. C. Alecio, M. Lyra and O. Rojas,
Spin frustration of a spin-1/2 Ising-Heisenberg three-leg tube as
an indispensable ground for thermal entanglement, J. Magn. Magn. Mats.
409 (2016) 124.

\bibitem{on-strk}O. Rojas, J. Stre\v{c}ka and S.M. de Souza, Thermal
entanglement and sharp specific-heat peak in an exactly solved spin-1/2
Ising-Heisenberg ladder with alternating Ising and Heisenberg inter--leg
couplings, Sol. Stat. Comm. 246 (2016) 68.

\bibitem{psd-Ising} J Stre\v{c}ka, Anomalous thermodynamic response
in the vicinity of pseudo-transition of a spin-1/2 Ising diamond chain,
Acta Phys. Pol. A 137 (2020) 610; arXiv:2002.06942.

\bibitem{pseudo}S. M. de Souza and O. Rojas, Quasi-phases and pseudo-transitions
in one-dimensional models with nearest neighbor interactions, Sol.
Stat. Comm. 269 (2018) 131.

\bibitem{Isaac}I. M. Carvalho, J. Torrico, S. M. de Souza, O. Rojas,
Oleg Derzhko, Correlation functions for a spin-12 Ising-XYZ diamond
chain: Further evidence for quasi-phases and pseudo-transitions, Ann.
Phys. 402 (2019) 45.

\bibitem{ph-bd}O. Rojas, A Conjecture on the Relationship Between
Critical Residual Entropy and Finite Temperature Pseudo-transitions
of One-dimensional Models, Braz. Jour. Phys. 50 (2020) 675; Residual
Entropy and Low Temperature Pseudo-Transition for One-Dimensional
Models, Acta Phys. Pol. A 137 (2020) 933.

\bibitem{unv-cr-exp}O. Rojas, J. Stre\v{c}ka, M. L. Lyra, S. M. de
Souza, Universality and quasicritical exponents of one-dimensional
models displaying a quasitransition at finite temperatures, Phys.
Rev. E 99 (2019) 042117.

\bibitem{Dec-trnsf} I. Syozi, Phase Transitions and Critical Phenomena,
Vol. 1, eds. C. Domb, M. S. Green, Academic Press, London, (1972)
269; M. Fisher, Transformations of Ising models, Phys. Rev. 113 (1959)
969; O. Rojas, J. S. Valverde, S. M. de Souza, Generalized transformation
for decorated spin models, Physica A 388 (2009) 1419; J. Stre\v{c}ka,
Generalized algebraic transformations and exactly solvable classical-quantum
models, Phys. Lett. A 374 (2010) 3718; O. Rojas, S. M. de Souza, Direct
algebraic mapping transformation for decorated spin models, J. Phys.
A: Math. Theor. 44 (2011) 245001.

\bibitem{Hutak} T. Hutak, T. Krokhmalskii, O. Rojas, S. M. de Souza,
O. Derzhko, Low-temperature thermodynamics of the two-leg ladder Ising
model with trimer rungs: A mystery explained, Phys. Lett. A, 387 (2021)
127020.

\bibitem{Weiguo}Weiguo Yin, Frustration-driven unconventional phase
transitions at finite temperature in a one-dimensional ladder Ising
model, arXiv:2006.08921; Finding and classifying an infinite number
of cases of the marginal phase transition in one-dimensional Ising
models, arXiv:2006.15087.

\bibitem{Tsai}Y. -C. Tsai, C. -K. Hu, Generalized antiferromagnetic
Heisenberg spin ladders, Physica B 305 (2001) 21.

\bibitem{R. Chen}R. Chen, H. Ju, H.-C. Jiang, O. A. Starykh, L. Balents,
Ground states of spin- triangular antiferromagnets in a magnetic field,
Phys. Rev. B 87 (2013) 165123.

\bibitem{Zhao}N. Zhao, H. ding, J. Zhang and Y. He, A low-energy
physics of an extended Hubbard chain with additional three-body couplings,
Chin. J. Phys. 56 (2018) 1633.

\bibitem{Zhang}H. Ding and J. Zhang, Metal-insulator transition in
an one-dimensional extended Hubbard model at quarter filling, Chin.
J. Phys. 54 (2016) 237.

\bibitem{Ding}X. Ma, H. Ding, Frustration-driven singlet superconductivity
in the one-dimensional model with positive interactions, Chin. J.
Phys. 55 (2017) 1888.

\bibitem{Shi}X. Shi, H. Ding, J. Zhang, Density wave instabilities
in the one-dimensional metals, Chin. J. Phys. 59 (2019) 250.

\bibitem{kramers}H. A. Kramers, G. H. Wannier, Statistics of the
Two-Dimensional Ferromagnet. Part I, Phys. Rev. 60 (1941) 252; Statistics
of the Two-Dimensional Ferromagnet. Part II, Phys. Rev. 60 (1941)
263.

\bibitem{ky lin}K. Y. Lin, An Elementary Proof of the Perron-Frobgenius
Theorem for Non-Negative Symmetric Matrices, Chin. J. Phys. 15 (1977)
283.

\bibitem{fritz}S. Buhrandt and L. Fritz, Antiferromagnetic Ising
model on the swedenborgite lattice, Phys. Rev. B 90 (2014) 094415.

\bibitem{otsuka} A.Otsuka, D. V. Konarev, R. N. Lyubovskaya, S. S.
Khasanov, M. Maesato, Y. Yoshida and G. Saito, Design of Spin-Frustrated
Monomer-Type C60\textbullet \textminus{} Mott Insulator, Crystals
8 (2018) 115.

\bibitem{mambrini}M. Mambrini, J. Trebosc and F. Mila, Residual entropy
and spin gap in a one-dimensional frustrated antiferromagnet, Phys.
Rev. B 59 (1999) 13806.

\bibitem{roj-alc}O. Rojas, F. C. Alcaraz, Phase diagram of a coupled
tetrahedral Heisenberg model, Phys. Rev. B 67 (2003) 174401.

\bibitem{vadim-1}V. Ohanyan, Antiferromagnetic sawtooth chain with
Heisenberg and Ising bonds, Cond. Matt. Phys., 12 (2009) 343.

\bibitem{Vadim-2}D. Antonosyan, S. Bellucci, V. Ohanyan, Exactly
solvable Ising-Heisenberg chain with triangular -Heisenberg plaquettes,
Phys. Rev. B 79 (2009) 014432.

\bibitem{Betchelor07}M.T. Batchelor, X.W. Guan, N. Oelkers, Z. Tsuboi,
Integrable models and quantum spin ladders: comparison between theory
and experiment for the strong coupling ladder compounds, Adv. Phys.
56 (2007) 465. 

\bibitem{Chiari}B. Chiari, O. Piovesana, T. Tarantelli, P.F. Zanazzi,
Exchange interaction in multinuclear transition metal complexes. 14.
Exchange interactions in a novel copper(II) linear-chain compound
with ladderlike structure: Cu2(1,4-diazacycloheptane)2Cl4, Inorg.
Chem. 29 (1990) 1172. 

\bibitem{Hiroi}Z. Hiroi, M. Azuma, M. Takano, and Y. Bando, A new
homologous series $\mathrm{Sr_{n-1}Cu_{n+1}O_{2n}}$ found in the
SrOCuO system treated under high pressure, J. Solid State Chem. 95
(1991) 230.

\bibitem{Willett}R. D. Willett, C. Galeriu, C. P. Landee, M. M. Turnbull,
B. Twamley, Structure and Magnetism of a Spin Ladder System:\LyXThinSpace{}
$\mathrm{(C_{5}H_{9}NH_{3})_{2}CuBr_{4}}$, Inorg. Chem. 43 (2004)
3804.

\bibitem{Onoda}M. Onoda, N. Nishiguci, Letter to the editor: crystal
structure and spin gap state of $\mathrm{CaV_{2}O_{5}}$, J. Solid
State Chem. 127 (1996) 359.

\bibitem{barnes} T. Barnes, J. Riera, Susceptibility and excitation
spectrum of $\mathrm{(VO)_{2}P_{2}O_{7}}$ in ladder and dimer-chain
models, Phys. Rev. B 50 (1994) 6817.

\bibitem{Simutis}G. Simutis, S. Gvasaliya, F. Xiao, C. P. Landee,
A. Zheludev, Raman study of spin excitations in the tunable quantum
spin ladder $\mathrm{Cu(Qnx)(Cl_{1-x}Br_{x})_{2}}$, Phys. Rev. B
93 (2016) 094412.
\end{thebibliography}
\end{document}